\RequirePackage{fix-cm}
\documentclass[twocolumn,epjc3]{svjour3}  
\smartqed  % flush right qed marks, e.g. at end of proof
\RequirePackage{graphicx}
\journalname{Eur. Phys. J. C}
\begin{document}
%
%\title{Insert your title here}
\title{Background model for the NaI(Tl) crystals in COSINE-100}



\author{P.~Adhikari\thanksref{addr1} 
	\and
	G.~Adhikari\thanksref{addr1} 
	\and
	E.~Barbosa de Souza\thanksref{addr2}
	\and
	N.~Carlin\thanksref{addr3}
	\and
	S.~Choi\thanksref{addr4}
	\and
	W.Q.~Choi\thanksref{addr5}
	\and
	M.~Djamal\thanksref{addr6}
	\and
	A.C.~Ezeribe\thanksref{addr7}
	\and
	C.~Ha\thanksref{addr8}
	\and
	I.S.~Hahn\thanksref{addr9}
	\and
	A.J.F.~Hubbard\thanksref{addr2}
	\and
	E.J.~Jeon\thanksref{corrauthor,addr8}
	\and
	J.H.~Jo\thanksref{addr2}
	\and
	H.W.~Joo\thanksref{addr4}
	\and
	W.G.~Kang\thanksref{addr8}
	\and
	M.~Kauer\thanksref{addr11}
	\and
	W.S.~Kang\thanksref{addr10}
	\and
	B.H.~Kim\thanksref{addr8}
	\and
	H.~Kim\thanksref{addr8}
	\and
	H.J.~Kim\thanksref{addr12}
	\and
	K.W.~Kim\thanksref{addr8}
	\and
	M.C.~Kim\thanksref{addr10}
	\and
	N.Y.~Kim\thanksref{addr8}
	\and
	S.K.~Kim\thanksref{addr4}
	\and
	Y.D.~Kim\thanksref{addr8,addr1}
	\and
	Y.H.~Kim\thanksref{addr8,addr13}
	\and
	V.A.~Kudryavtsev\thanksref{addr7}
	\and
	H.S.~Lee\thanksref{addr8}
	\and
	J.~Lee\thanksref{addr8}
	\and
	J.Y.~Lee\thanksref{addr4}
	\and
	M.H.~Lee\thanksref{addr8}
	\and
	D.S.~Leonard\thanksref{addr8}
	\and
	W.A.~Lynch\thanksref{addr7}
	\and
	R.H.~Maruyama\thanksref{addr2}
	\and
	F.~Mouton\thanksref{addr7}
	\and
	S.L.~Olsen\thanksref{addr8}
	\and
	H.K.~Park\thanksref{addr15}
	\and
	H.S.~Park\thanksref{addr13}
	\and
	J.S.~Park\thanksref{addr8}
	\and
	K.S.~Park\thanksref{addr8}
	\and
	W.~Pettus\thanksref{addr2}
	\and
	H.~Prihtiadi\thanksref{addr6}
	\and
	S.~Ra\thanksref{addr8}
	\and
	C.~Rott\thanksref{addr10}
	\and
	A.~Scarff\thanksref{addr7}
	\and
	N.J.C.~Spooner\thanksref{addr7}
	\and
	W.G.~Thompson\thanksref{addr2}
	\and
	L.~Yang\thanksref{addr14}
	\and
	S.H.~Yong\thanksref{addr8}	
}

\thankstext{corrauthor}{Corresponding author: ejjeon@ibs.re.kr}


\institute{Department of Physics and Astronomy, Sejong University, Seoul 05006, Korea \label{addr1}
	\and
	Wright Laboratory, Department of Physics, Yale University, New Haven, CT 06520, USA \label{addr2} 
	\and
	Physics Institute, University of S\~{a}o Paulo, 05508-090, S\~{a}o Paulo, Brazil \label{addr3}
	\and
	Department of Physics and Astronomy, Seoul National University, Seoul 08826, Republic of Korea \label{addr4}
	\and
	Korea Institue of Science and Technology Information, Daejeon, 34141, Republic of Korea \label{addr5}
	\and
	Department of Physics, Bandung Institute of Technology, Bandung 40132, Indonesia \label{addr6}
	\and
	Department of Physics and Astronomy, University of Sheffield, Sheffield S3 7RH, United Kingdom \label{addr7}
	\and
	Center for Underground Physics, Institute for Basic Science (IBS), Daejeon 34126, Republic of Korea \label{addr8}
	\and
	Department of Science Education, Ewha Womans University, Seoul 03760, Republic of Korea \label{addr9}
	\and
	Department of Physics, Sungkyunkwan University, Seoul 16419, Republic of Korea \label{addr10}
	\and
	Department of Physics and Wisconsin IceCube Particle Astrophysics Center,
	University of Wisconsin-Madison, Madison, WI 53706, USA \label{addr11}
	\and
	Department of Physics, Kyungpook National University, Daegu 41566, Republic of Korea \label{addr12}
	\and
	Korea Research Institute of Standards and Science, Daejeon 34113, Republic of Korea \label{addr13}
	\and
	Department of Physics, University of Illinois at Urbana-Champaign, Urbana, IL 61801, USA \label{addr14}
	\and
	Department of Accelerator Science, Graduate School, Korea University, Sejong 30019, Korea \label{addr15}
           %\and
           %\emph{Present Address:} if needed\label{addr3}
}

\date{Received: date / Accepted: date}

\maketitle

\begin{abstract}
The COSINE-100 dark matter search experiment is an array of NaI(Tl) crystal detectors located in the Yangyang Underground Laboratory (Y2L).
To understand measured backgrounds in the NaI(Tl) crystals we have performed Monte Carlo simulations using the Geant4 toolkit and developed background models for each crystal that consider contributions from both internal and external sources, including cosmogenic nuclides.
The background models are based on comparisons of measurement data with Monte Carlo simulations that are guided by a campaign of material assays and are used to evaluate backgrounds and identify their sources. The average background level for the six crystals (70 kg total mass) that are studied is 3.5 counts/day/keV/kg in the (2--6) keV energy interval. The dominant contributors in this energy region are found to be $^{210}$Pb and $^3$H.
\end{abstract}  

\section{Introduction}
\label{intro}
COSINE-100 is a dark matter search experiment consisting of a 106 kg array of eight ultra-pure NaI(Tl) crystals~\cite{kims-nai2014,kims-nai2015}. Its primary goal is to test DAMA/LIBRA's assertion of an observation of annual modulation signal~\cite{bernabei08,bernabei10,bernabei13,bernabei18}. The experiment has been operating at the Yangyang Underground Laboratory (Y2L) since September 2016~\cite{cosinedet17}.
COSINE-100 is one of several NaI(Tl)-based dark matter searches in operation (DM-Ice17~\cite{dmice17}, ANAIS~\cite{amare14,amare16}) or under development (DM-Ice~\cite{dmice14}, Kam-LAND-PICO~\cite{kamland-pico}, SABRE~\cite{sabre17}, COSINUS~\cite{cosinus}). 
Previously, the KIMS-CsI experiment put a limit on interaction rates in CsI crystals~\cite{hslee07,sckim12} that precluded the interpretation of the DAMA modulation signal as being due to WIMPs scattering from I or Tl nuclei, considering the different quenching factors of iodine and thallium for NaI(Tl) and CsI(Tl).

The COSINE-100 crystal array is immersed in a tank of liquid scintillator (LS)
that tags backgrounds that originate from outside the LS as well as decays of  $^{40}$K nuclides inside the crystals. 
To determine the sources of the backgrounds, we have performed Monte Carlo simulations using the Geant4 toolkit (V.4.9.6.p02) ~\cite{geant4} and built a background model for the eight detectors by iteratively fitting their contributions to the measured energy spectra; %two crystals are excluded in this paper due to their low light yields and relatively higher background contamination. 
two crystals are excluded in this paper due to their low light yields, which result in a background spectrum without characteristic peaks of isotopes by the worse energy resolution, and relatively higher background contamination in the low energy region.

The paper is structured as follows: the COSINE-100 experimental setup is described in Sect.~\ref{sec:2}. Section~\ref{sec:3} describes the background modeling, with the simulation method described in Sect.~\ref{sec:3.2}, sources of the background internal and external to the crystal and of cosmogenic origin in Sects.~\ref{sec:3.2} --~\ref{sec:3.4}.  Section~\ref{sec:4} describes the comparison and fit to the data, and Sect.~\ref{sec:5} provides discussions of the background developed by the fits. Finally Sect.~\ref{conc} provides conclusions.

\section{COSINE-100 setup and simulation geometry}
\label{sec:2}

\begin{figure*}
\begin{center}
\begin{tabular}{cc}
\includegraphics[width=0.4\textwidth]{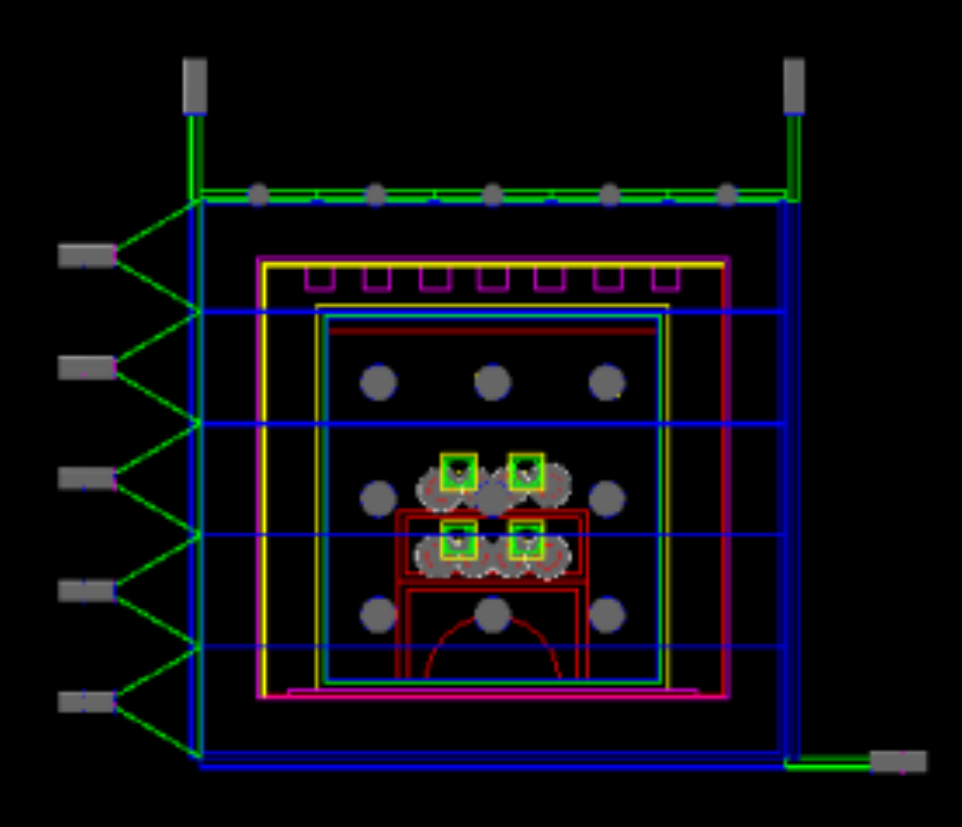} &
\includegraphics[width=0.47\textwidth]{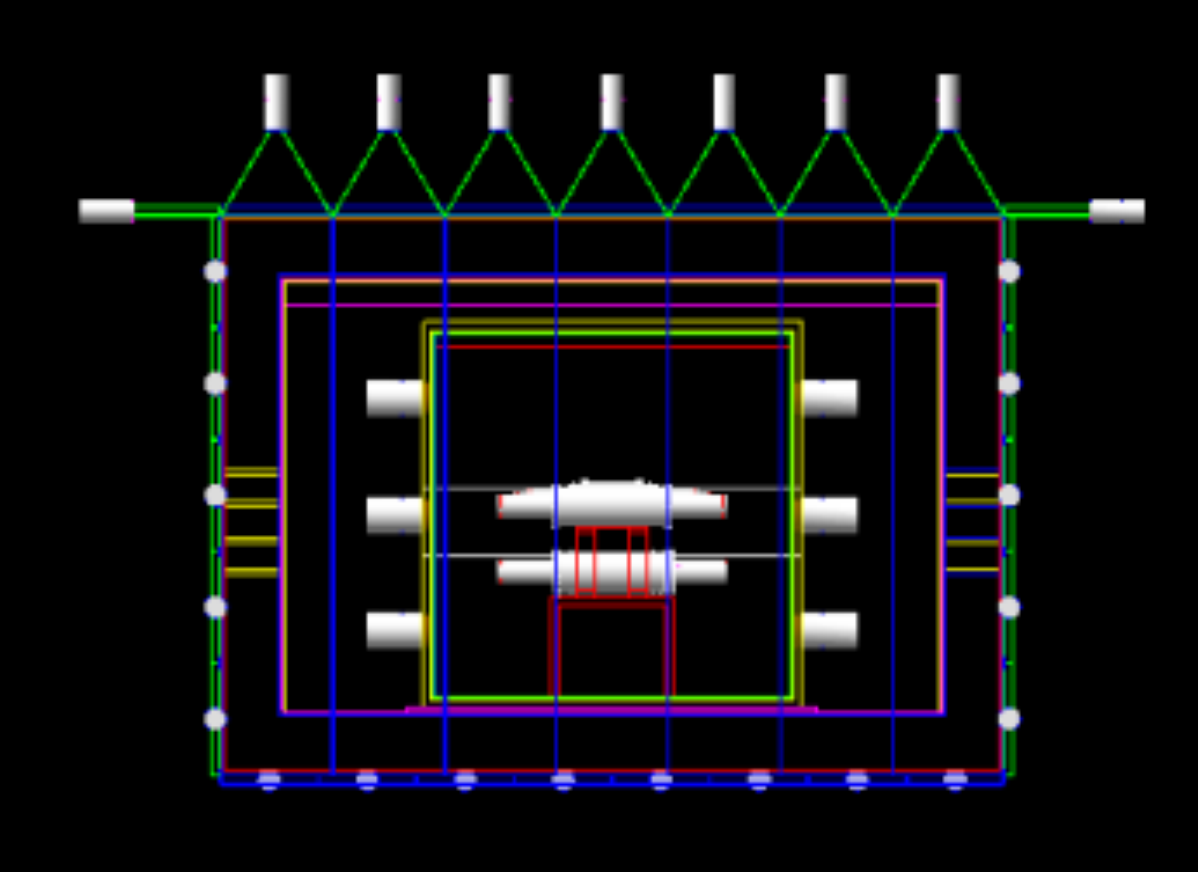} \\
(a) Front view  & (b) Side view  \\
\end{tabular}
\caption{
Front (a) and Side (b) views of the detector geometry used in the Geant4 simulations. From outside inward, the four shielding layers include: 
3 cm thick plastic scintillator panels (dark blue), 20 cm lead (pink), 3 cm copper box (light green), and liquid scintillator (not shown).
}
\label{detector-setup}
\end{center}
\end{figure*}

The experimental setup is described in detail in Ref.~\cite{cosinedet17}. The simplified geometry used for the simulations is shown in Fig.~\ref{detector-setup}.
Eight NaI(Tl) crystals, arranged in two layers, are located in the middle of a four-layer shielding structure. 
From outside inward, the four shielding layers are plastic scintillator panels, a lead-brick castle, a copper box, and a scintillating liquid. The eight NaI(Tl) crystal assemblies and their support table are immersed in the scintillating liquid that serves both as an active veto and a passive shield. 

The eight NaI(Tl) crystals were grown out of batches of powder provided by Alpha Spectra~\cite{as-inc} with successive improvements. The first attempts, which produced an order of magnitude reduction in $^{40}$K, were AS-B and AS-C. This was followed by WIMPScint-II (AS-WSII) which reduced the $^{210}$Pb contamination, and WIMPScint-III (AS-WSIII) which resulted in another factor of two reduction of $^{40}$K. The results are summarized in Table~\ref{measured-activity}.
The final crystals are cylindrically shaped and hermetically encased in OFE copper tubes with wall thickness of 1.5 mm and quartz windows (12.0 mm thick) at each end.
Each crystal's lateral surfaces were wrapped in roughly 10 layers of 250~$\mu$m-thick PTFE reflective sheets. 
The quartz windows are optically coupled to each end of the crystal via 1.5~mm thick optical pads. These, in turn, are optically coupled to 3-inch Hamamatsu R12669SEL 
photomultiplier tubes (PMTs) with a thin layer of high viscosity optical gel.
The PMTs are protected from the liquid scintillator by a housing made of copper and PTFE.

The following components of the detector have been included in the simulation: PTFE reflective sheets, copper tubes, quartz windows, optical gel, PMT housing, and PMTs.

\section{Background modeling}
\label{sec:3}

\subsection{Simulation method}
\label{sec:3.1}
The Physics list classes of G4EmLivermorePhysics for low energy electromagnetic process and G4Radioactive-Decay for radioactive decay process were used~\cite{geant4:lowEM,geant4:lowEMpackage,geant4:2016}.
The $^{238}$U and $^{232}$Th decay chains were treated as broken at the long-lived parts of the chain.  
The $^{238}$U chain was broken into five distinct groups and the $^{232}$Th chain was broken into three groups. The details are reported in Ref.~\cite{kims-nai-bkg17}.

Each simulated event record includes all energy deposited in the crystals within an event window of 10~$\mu$s from the time a decay is generated, to account for the conditions in the data acquisition system (DAQ) of the experimental setup~\cite{cosinedet17}.  
Sometimes decays with relatively short half-lives, such as $^{212}$Po decay (with a half-life of 300~ns) and the subsequent daughter decays will appear in the 10~$\mu$s time window, resulting in pileup events. They are treated as a single event in the simulation.

The simulated spectrum was convolved with an energy dependent energy resolution function developed during a calibration run.
Calibration points were measured using $\gamma$--ray sources: 59.5~keV($^{241}$Am), 1173.2~keV and 1332.5~keV ($^{60}$Co).
Internal background peaks at 3.2 and 1460.8 keV from $^{40}$K, 67.3~keV from $^{125}$I, and 609.3~keV from $^{214}$Bi were used to calibrate the measured spectra; peaks at 3.2~keV, 59.5~keV, and 67.3~keV were used for the low energy calibration below 70~keV.   
  
\subsection{Internal backgrounds in the NaI(Tl) crystals}
\label{sec:3.2}

\begin{table*}[ht]
\begin{center}
  \caption{
Radioactive contamination levels in the COSINE-100 crystals reported in Ref.~\cite{cosinedet17}. These were used as inputs when fitting the simulations to the data.
The $^{238}$U and $^{232}$Th decays were assumed to be in equilibrium.
  }
\label{measured-activity}
\begin{tabular}{lccccccc}
  \hline                                                                                
  Crystal    & Mass & Size (inches)     & Powder    & $\alpha$ Rate & $^{40}$K  & $^{238}$U & $^{228}$Th        \\
             & (kg) & (diameter$\times$length)     &           & (mBq/kg) & (ppb)  & (ppt)  & (ppt)           \\
  \hline                 
  Crystal-1  & 8.3  & $5.0\times7.0$     &  AS-B     & $3.20\pm0.08$  & $34.7\pm4.7$  & \textless0.02  & $1.3\pm0.4$    \\
  Crystal-2  & 9.2  & $4.2\times11.0$     &  AS-C     & $2.06\pm0.06$ & $60.6\pm4.7$   & \textless0.12  & \textless0.6    \\
  Crystal-3  & 9.2  & $4.2\times11.0$     &  AS-WSII  & $0.76\pm0.02$ & $34.3\pm3.1$    & \textless0.04  & $0.4\pm0.2$     \\
  Crystal-4  & 18.0 & $5.0\times15.3$     &  AS-WSII  & $0.74\pm0.02 $ &$33.3\pm3.5$    &                & \textless0.3       \\
  Crystal-5  & 18.3 & $5.0\times15.5$     &  AS-C     & $2.06\pm0.05 $ & $82.3\pm5.5$  &                & $2.4\pm0.3$     \\
  Crystal-6  & 12.5 & $4.8\times11.8$     &  AS-WSIII & $1.52\pm0.04 $ & $16.8\pm2.5$   & \textless0.02  & $0.6\pm0.2$      \\
  Crystal-7  & 12.5 & $4.8\times11.8$     &  AS-WSIII & $1.54\pm0.04 $ & $18.7\pm2.8$   &                & \textless0.6    \\
  Crystal-8  & 18.3 & $5.0\times15.5$     &  AS-C     & $2.05\pm0.05  $  & $54.3\pm3.8$ &                & \textless1.4     \\
   \hline  
\end{tabular}
\end{center}
\end{table*}

After the insertion of the crystals into the shield and prior to filling the liquid scintillator container, their background levels were
measured to verify that they were free of any additional contamination.
Overall, the eight crystals have acceptable $^{238}$U and $^{232}$Th contaminations as shown in Table~\ref{measured-activity}~\cite{cosinedet17}.
%Chain equilibrium 
Secular equilibrium in the chains is assumed for the interpretation of $^{238}$U and $^{232}$Th
related radioactivity measurements, with the exception of $^{210}$Pb.

In order to estimate the background contributions from $^{238}$U, $^{232}$Th, $^{40}$K, and $^{210}$Pb, 
we simulated background spectra from the internal radioactive contaminants and normalized them by their measured activities in Table~\ref{measured-activity}.
In the normalization we assumed a chain equilibrium and, thus, all related activities within the chains are equal to the $^{238}$U, $^{232}$Th, and $^{40}$K activities multiplied by the branching ratios for decay of the daughter isotopes.   
We also added the background simulation of internal $^{210}$Pb by considering the measured $\alpha$ rate. 
The resultant background contributions, except for those from $^{40}$K and $^{210}$Pb, were negligible in all eight crystals.

The $^{40}$K contribution is reduced by the LS veto detector.
To measure the reduction efficiency of the $^{40}$K generated 3.2~keV emission background provided by
tagging the accompanying 1460.8~keV $\gamma$-ray in one of the other NaI(Tl) crystals or
the LS, and to compare this to the efficiency provided by the other crystals alone,
we generated $^{40}$K decays at random locations inside a NaI(Tl) crystal for the
cases with and without the LS veto.
From these simulations, we determined that
the Crystal-6 tagging efficiency by other crystals without LS is 31.7$\pm$0.1\,$\,\%$ and
by the LS only is 64.9$\pm$0.2$\,\%$. The total combined efficiency is 81.7$\pm$0.3\,$\%$.
The efficiency is measured in the crystal energy range between 2 and 6\,keV
by requiring the LS energy deposit be larger than 20\,keV.
Efficiencies vary depending on the crystal location in the detector.
For example, Crystal-1 (at the corner of the 4$\times$2 array) shows higher coverage by the LS (75\,\%) than
neighboring crystals (17\,\%), but the combined efficiency is similar to that of Crystal-6 (82\,\%).
The tagging efficiency of the 1460.8\,keV $\gamma$-ray in the LS-only case is lower because
the range of the $\gamma$-ray in the NaI(Tl) crystal is shorter than in the LS.
Therefore, more $\gamma$-rays are stopped in the other crystals than in the LS.
These estimated efficiencies are in agreement with measurements~\cite{cosinedet17}.
Accordingly, the $^{40}$K background level is reduced by as much as 80\,\% by requiring single-hit crystal events with no signal in the LS. 

The $^{210}$Pb contribution is estimated by modeling the background from bulk $^{210}$Pb and surface $^{210}$Pb as discussed in Sect.~\ref{sec:4}.
    
\subsection{External background sources}
\label{sec:3.3}

\begin{table}
\begin{center}
\caption{
Radioactivity levels in detector components inside the shielding.
(a) The radioactivities were measured with a HPGe detector at Y2L; upper limits are quoted with 90\% C.L. The PMTs are measured in units of mBq/PMT and the other external sources are measured in units of mBq/kg
(b) SEL means ``selected for high quantum efficiency''.
}
\label{photomultipliertubes}
\begin{tabular}{cccc}
\hline
 & \multicolumn{3}{c}{Radioactivity$^{a}$} \\
 External source & U($^{214}$Bi) & Th($^{228}$Ac) & ($^{40}$K) \\ \hline
 PMT~\cite{kims-nai2014} & 25 $\pm$ 5 &  12 $\pm$ 5 &  58 $\pm$ 5   \\ 
 (R12669SEL$^{b}$) & & & \\
Quartz window & $<$1.8 &  $<$7.5 &  $<$20   \\ 
PTFE reflector  & $<$0.5 &  $<$1.0 &  $<$6.4   \\ 
Cable ties & $<$4.2 &  $<$3.5 &  149 $\pm$ 32   \\ 
LS  & $<$2.7 &  $<$3.3 &  7 $\pm$ 4   \\ 
\hline
\end{tabular}
\end{center}
\end{table} 

The external $\gamma$ background from the radioactive isotopes in the surrounding rocks is shielded by the 20~cm-thick lead castle and the 3~cm-thick copper box. By using the full shielding structure with $N_{2}$ gas flowing into the inside of the copper shield to avoid backgrounds from $^{222}$Rn in the air at Y2L (measured to be $1.20\pm 0.49$ pCi/L~\cite{kims-radon2011}), we reduced the environmental background by a factor of 10,000 based on the measurements of a high-purity Ge (HPGe) detector, thus ensuring that those contributions are negligibly small.

Despite all the efforts to block backgrounds due to external sources, 
some backgrounds from radioactive contaminations in detector components inside the shielding are still expected, including from the PMTs, grease, copper case, bolts, cables, acrylic supports, liquid scintillator, copper box, and steel that supports the lead block housing.
We simulated background spectra from those external sources to test their effects and compared the shapes of contributions to the crystals' energy spectra. 
We found that all the spectra from these external sources are similar in shape and, thus, could be represented by a spectrum that is obtained by simulating $^{238}$U, $^{232}$Th, and $^{40}$K, distributed randomly in the volume outside the eight crystals.
Because the PMTs are the main contributer to the external background we used two kinds of spectra for the external background modeling; one is the spectrum from the PMTs and another is the spectrum from the other external sources that is treated as a parameter floating in the fit.   
The radioactivity levels of the PMTs and PMT surrounding parts were measured underground with a HPGe detector and the results are listed in Table~\ref{photomultipliertubes}.
We used the measured activities from the PMTs to constrain the data fitting and treated background contributions from the PMTs in nine groups as broken at the long-lived parts of the chain.

\subsection{Treatment of cosmogenic radionuclides}
\label{sec:3.4}

\begin{table*}
\begin{center}
\caption{Cosmogenic radionuclides in NaI(Tl) crystal (a) and exposure time and radioactivity cooling time at Y2L (b).
}
\label{cosmogenic}
\begin{minipage}{0.47\textwidth}
\begin{tabular}{c|c|c}
\multicolumn{3}{c}{(a)} \\ \hline
Cosmogenic & Half-life & Production rate \\
isotopes & (days) & at sea level~\cite{walter-thesis} \\
 & & (counts/kg/day) \\ \hline
$^{125}$I & 59.4 & 221 \\
$^{121}$Te & 19.17 & 93 \\
$^{121m}$Te & 164.2 & 93 \\
$^{123m}$Te & 119.2 & 52 \\
$^{125m}$Te & 57.4 & 74 \\
$^{127m}$Te & 106.1 & 93 \\ 
$^{113}$Sn & 115.1 & 9.0 \\ \hline
$^{109}$Cd & 461.4 & 4.8 \\
$^{3}$H & 4500 & 26 \\
$^{22}$Na & 951 & 66 \\ \hline
\end{tabular}
\centering
\end{minipage}
\hfill
\begin{minipage}{0.47\textwidth}
\begin{tabular}{c|c|c}
\multicolumn{3}{c}{(b)} \\ \hline
Crystal & Exposure & Radioactivity \\ 
&  time (see text) & cooling time at Y2L \\
 & (years) & (years) \\ \hline
Crystal-1 & 2 & 3 \\
Crystal-2 & 0.75 & 2.75 \\
Crystal-3 & & 1.2 \\
Crystal-4 & 1.7 & 0.5 \\
Crystal-6 & 0.3 & 0.6 \\
Crystal-7 & 0.3 & 0.6 \\ \hline
\end{tabular}
\end{minipage}
\end{center}
\end{table*}
 
Although the eight NaI(Tl) crystals had underground radioactivity cooling times that ranged from several months to three years, there are still background contributions due to the long-lived cosmogenic isotopes that were activated by cosmic rays while they were on the surface.

To consider these backgrounds, we first checked the list of cosmogenic radioactive isotopes that are produced in NaI(Tl), as reported in Ref.~\cite{walter-thesis,cosmogenic-amre15,cosmogenic-villar18,cosmogenic-amre18}.
In Table~\ref{cosmogenic}~(a), we list the contributing cosmogenic isotopes with their half lives; short-lived isotopes, for which half lives are less than a year, are $^{125}$I, $^{121}$Te, $^{121m}$Te, $^{123m}$Te, $^{125m}$Te, $^{127m}$Te, and $^{113}$Sn and long-lived isotopes are $^{109}$Cd, $^{3}$H, and $^{22}$Na. 
The radioactivity cooling time at Y2L for each crystal at the time data-taking for COSINE-100 started, is listed in Table~\ref{cosmogenic}~(b).    
The short-lived isotopes are not expected to contribute to either Crystal-1 or Crystal-2 because their cooling times are long enough to reduce these activities to a %low level.
negligible level. 

However, we expect some backgrounds from the sho-rt-lived isotopes in other crystals because their production rates at sea level, as listed in Table~\ref{cosmogenic}~(a), are high and their cooling times are less than or equal to a year. 
In addition, there are long-lived $^{109}$Cd, $^{3}$H, and $^{22}$Na nuclides that are potentially hazardous background sources; ${\it e.g.}$, the beta-decay spectrum of tritium has an endpoint energy of 18 keV. We thus need to understand their background contributions in the low energy region, especially in the (2--6)~keV WIMP signal region of interest (ROI). 
Because it is impossible to compute the initial activities of those isotopes from the production rates in each crystal at Y2L without knowing the cosmic ray exposure conditions: time, location, altitude, etc.~\cite{walter-thesis},
we investigated the correlation of characteristic peaks produced by $\gamma$/X-rays from the decay of cosmogenic isotopes.

\begin{itemize}
  \item[$\bullet$] 
  %$^{109}$Cd decays to the isomeric state of $^{109}$Ag emitting X-rays of 22$\sim$25~keV, by L-shell electron capture
  $^{109}$Cd decays by electron capture to the isomeric state of $^{109}$Ag depositing in the crystal the binding energy of the Ag K-shell electrons (25.5~keV), that will be accompanied by the 88~keV $\gamma$ ray from the isomer transition of $^{109}$Ag having a mean time of 57.4 seconds. %after 57.4 seconds of the mean time.
By using the timing information of two adjacent events that have each 25.5~keV and 88~keV, we measured the background contribution of $^{109}$Cd in Crystal-4 and found it to be 0.10$\pm$0.01~mBq/kg.  \\%the NaI(Tl) crystal. \\
  \item[$\bullet$] 
  $^{22}$Na decays via positron emission (90\%) and electron capture (10\%), followed by 1274.6~keV $\gamma$-ray emission with a mean lifetime of 3.8 yr. 
The electron capture decay produces 0.9~keV emissions. Therefore, $\sim$10\% of the $^{22}$Na decay will produce 0.9~keV X-rays and 1274.6~keV $\gamma$ rays simultaneously. Meanwhile, the positron will be converted to two 511~keV annihilation $\gamma$ rays. %If one of the two 511 and 1274.6~keV $\gamma$ rays escapes from a crystal, the energy deposited in the crystal will be 650--1000~keV.  We looked for a coincidence events that deposit 1274.6~keV of energy in one NaI(Tl) crystal and a %$\gamma$-ray hit
%$\beta$+ and a 511~keV gamma emission in the 650 to 1000-keV energy interval in another NaI(Tl) crystal.
\end{itemize}

However, it is generally difficult to measure long-lived cosmogenics' activities, such as those for $^{3}$H, directly from the data due to their long half-lives.
Therefore, we simulated background spectra from cosmogenic isotopes listed in Table ~\ref{cosmogenic}~(a) and used their shapes in the data fitting, while floating their unknown fractions.
%The measured activities and MC-determined efficiencies can be used to constrain the fit.
The details of their treatment in the background model for each NaI(Tl) crystal are discussed in section~\ref{sec:4}.

\section{Comparison with measured data}
\label{sec:4}
To model the COSINE-100 detector backgrounds, we used data collected between Oct 21 and Dec 19 2016. We applied low energy noise cuts as described in Ref.~\cite{cosinedet17}. 
The LS veto threshold was set at 20 keV for both single crystal and multiple crystal events. Events in coincidence with an LS signal ($>$20 keV in LS) and/or more than 4 photo electrons in other crystals are defined as multiple events.

Crystals 1 and 2 have no short-lived cosmogenic contaminants and were used for comparisons.
Since Crystal-3 and Crystal-4 have different production times and delivery times at the Y2L, their expected short-lived cosmogenic activities are quite different. But they are made of the same NaI powder and expected to have similar internal activities.
Crystal-6 and Crystal-7 are twin crystals made up of the same NaI powder and at the same time. They are installed between Crystal-5 and 8 and expected to have similar external background.
Both crystals are expected to have the same amount of internal, external and cosmogenic activities. We compare the activities of these crystal subdivisions.

We use a log likelihood method to fit the data. The fitting range is 6 keV--2MeV and we perform four  simultaneous fits: single-hit low-energy, single-hit high-energy, multiple-hit low-energy and multiple-hit high-energy. Because different FADC systems are used for low- and high-energy data the resolutions are different. Low-energy means below 70 keV and high-energy means above 70 keV.  
The internal $^{238}$U, $^{232}$Th, and $^{40}$K levels are constrained to be within 20\% of their measured values.
We use a 10 $\mu$m thickness to distinguish between surface and bulk $\rm ^{210}Pb$ components to study surface contamination in the crystals that is generated uniformly within the thickness; their levels are allowed to float.  
The $^{238}$U, $^{232}$Th, and $^{40}$K levels in the PMTs are constrained to be within 50\% of their measurements, taking into account contributions from PMT surrounding parts.  Cosmogenic  and the external  $^{238}$U, $^{232}$Th, $^{40}$K, and  $^{60}$Co levels from other parts of the detector are free floated.

\subsection{Background model for crystals 1 and 2}
\label{sec:4.1}

\begin{figure*}
\begin{center}
\begin{tabular}{cc}
\includegraphics[width=0.5\textwidth]{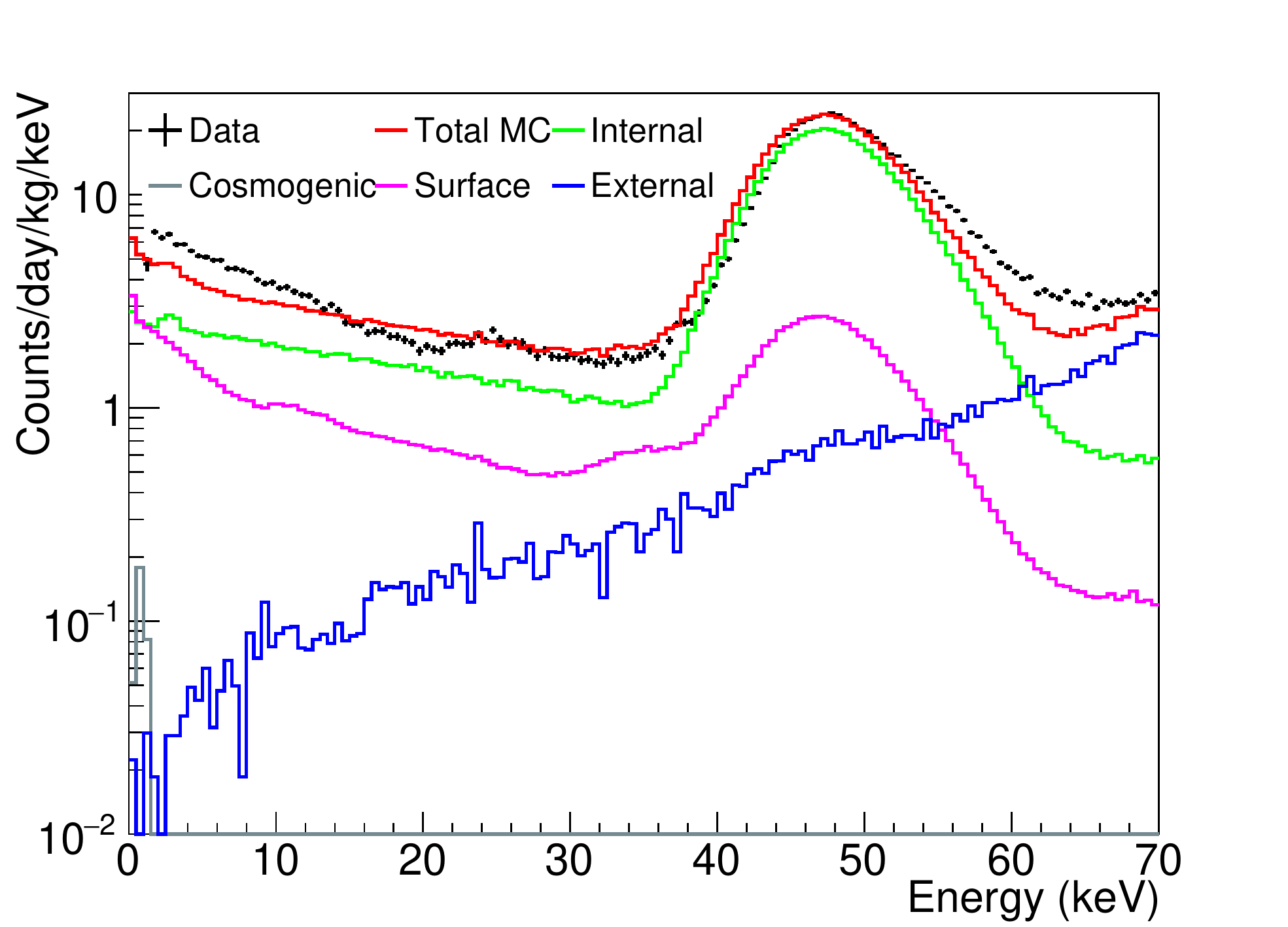} &
\includegraphics[width=0.5\textwidth]{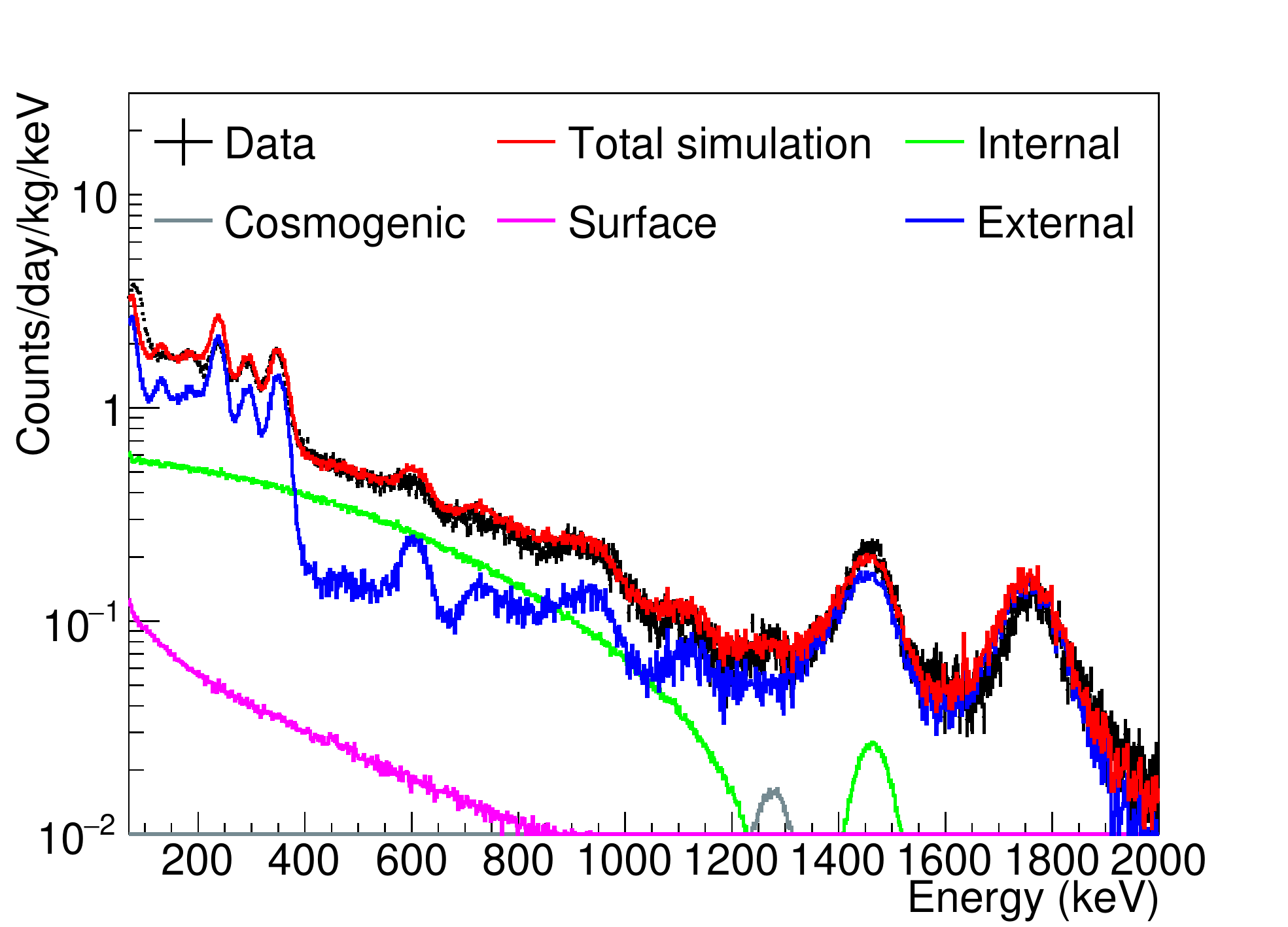} \\
(a) Single-hit events with low-energy data  & (b) Single-hit events with high-energy data\\
\includegraphics[width=0.5\textwidth]{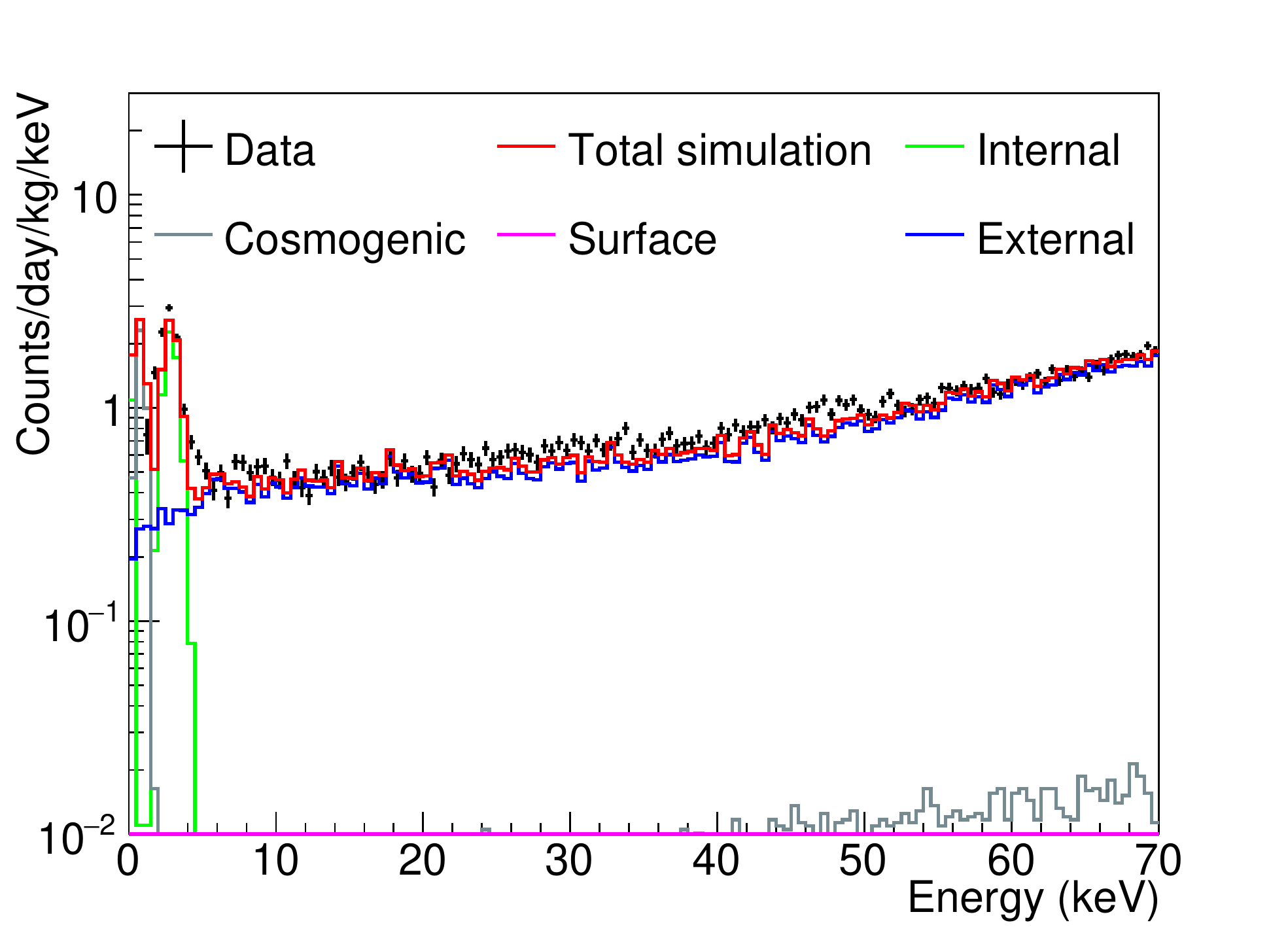} &
\includegraphics[width=0.5\textwidth]{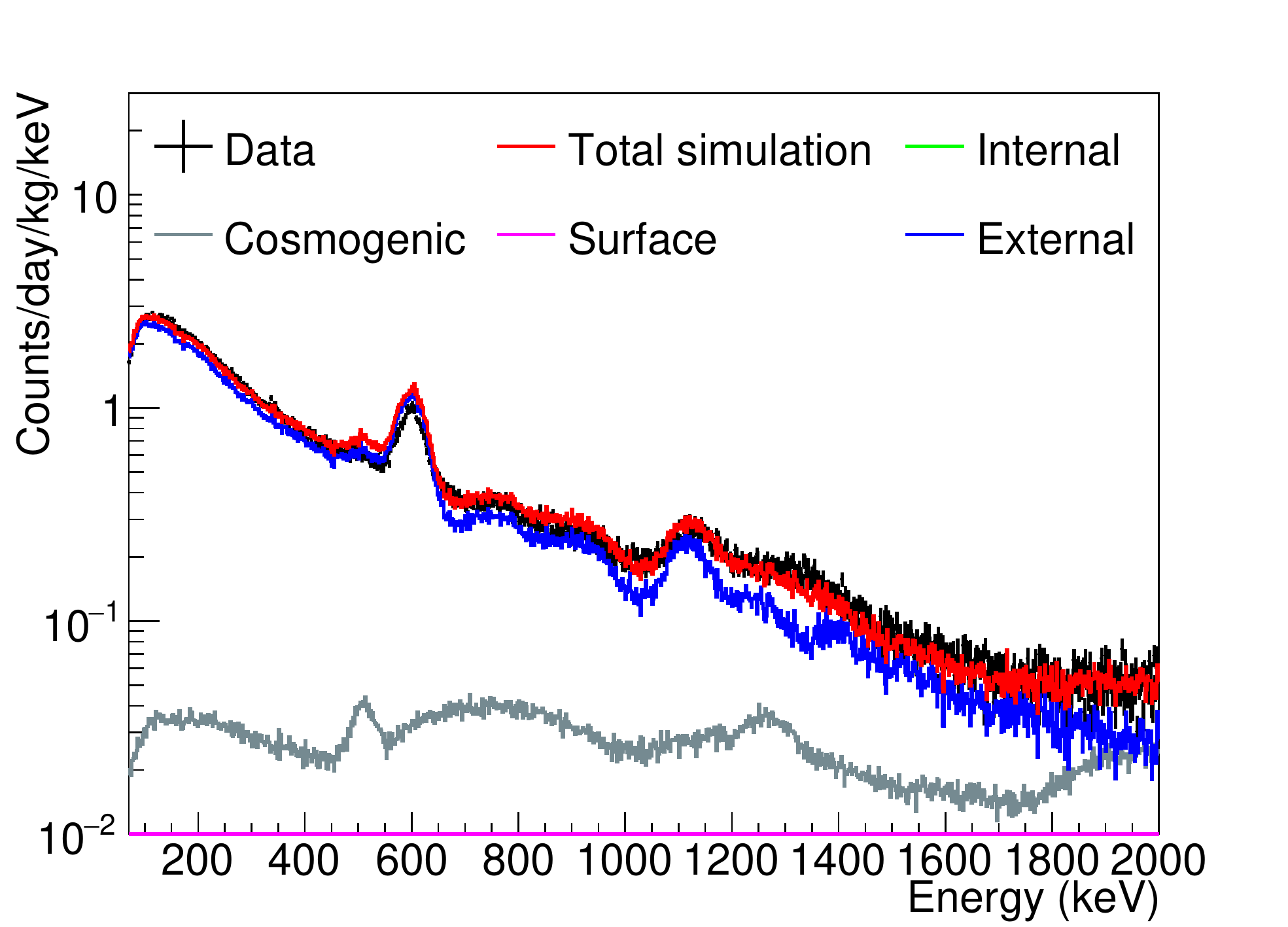} \\
(c) Multiple-hit events with low-energy data  & (d) Multiple-hit events with high-energy data\\
\end{tabular}
\caption[]{
Measured single- and multiple-hit background spectra of Crystal-1 fitted with all simulated background spectra using four channel simultaneous fitting.
$^{3}$H and $\rm ^{109}Cd$ are excluded in the fit.
}
\label{C1_fit_NoH3}
\end{center}
\end{figure*}

\begin{figure*}
\begin{center}
\begin{tabular}{cc}
\includegraphics[width=0.495\textwidth]{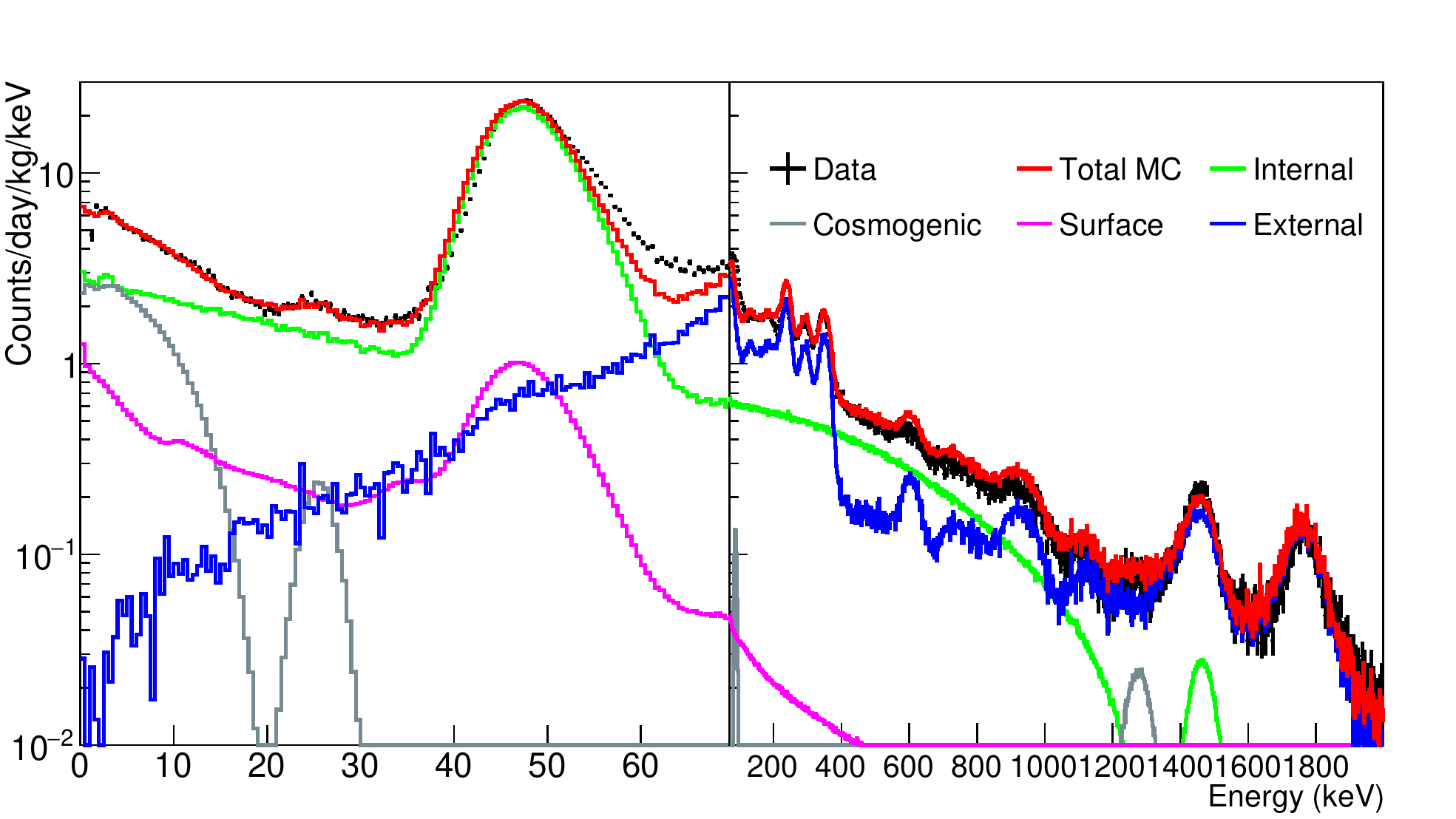} &
\includegraphics[width=0.495\textwidth]{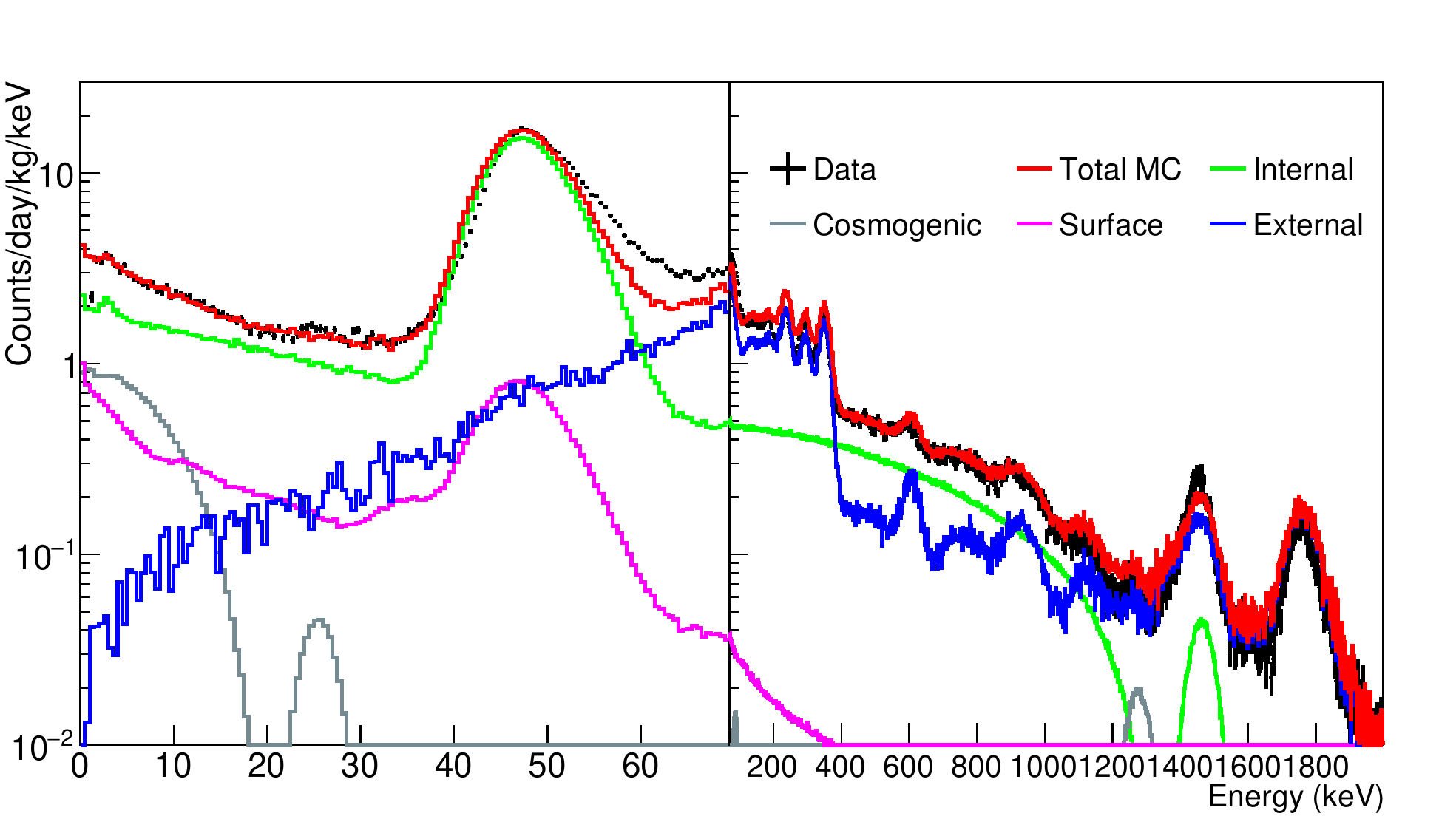} \\
(a)  Crystal-1 & (b) Crystal-2  \\
\end{tabular}
\caption[]{Measured single-hit background spectra of Crystal-1 and Crystal-2 fitted with all simulated background spectra.
$^{3}$H and $\rm ^{109}Cd$ are included in the fit. 
%The agreement between the measured and fitted multiple-hit background spectra of Crystal-2, 3, 4, 6, and 7 is as good as shown for Crystal-1.
}
\label{C1_fit}
\end{center}
\end{figure*}

The surface exposure time of Crystal-1 is longer than Crystal-2 by more than a year and it is expected to have more long-lived cosmogenic isotopes such as $^{3}$H and $\rm ^{109}Cd$. 
In addition, it was delivered by air while Crystal-2 was shipped by sea.  
At first we did not consider long-lived cosmogenic isotopes in the data fitting for Crystal-1 and the resulting four-channel fits are shown in Fig.~\ref{C1_fit_NoH3}. The overall energy spectrum is well matched to the data for both single-hit and multiple-hit events, except for the single-hit low-energy events. 
The agreement between the measured and fitted multiple-hit background spectra of Crystal-2, 3, 4, 6, and 7 is as good as shown for Crystal-1.

The peak around 46~keV is due to $^{210}$Pb in the crystals. 
However only a small part, 4.3\%, is due to the 46.5~keV gamma-ray line;  most of the events in the peak are from the conversion electrons, Auger electrons, and X-rays, followed by beta electrons from the decay to $^{210}$Bi. As a result, the peak is non-Gaussian. It is not well reproduced by the simulations using the Geant4 version 4.9.6.p02 and, thus, we will consider a higher version number for further modeling. 

We found that it was not possible to model the Crystal-1 and 2 background spectra for energies below 30 keV with only bulk and surface $^{210}$Pb contaminations (see Fig.~\ref{C1_fit_NoH3}(a)).  To get adequate fits to both crystals, we had to include significant contributions from $^{3}$H and $^{109}$Cd, as shown in Fig.~\ref{C1_fit}(a) and~\ref{C1_fit}(b). They are also included in the models for the other crystals. 

Internal $^{210}$Pb contamination levels independently determined from the alpha activities in %Crystal-1 and 2 
the crystals are listed in Table~\ref{measured-activity}. To study surface contamination in the crystals both bulk and surface $^{210}$Pb components are free floated in the fit. 
The $^{40}$K contamination levels in the %two crystals
crystals are identified by coincident signals between a 3.2 keV emission in one NaI(Tl) detector and a 1460.8~keV gamma-ray in one of the other NaI(Tl) crystals or an energy deposition in the LS. 
Background from the readout PMTs and surrounding material is considered as external components. 

\subsection{Background model for crystals 3 and 4}
\label{sec:4.2}

\begin{figure}
\begin{center}
\includegraphics[width=0.5\textwidth]{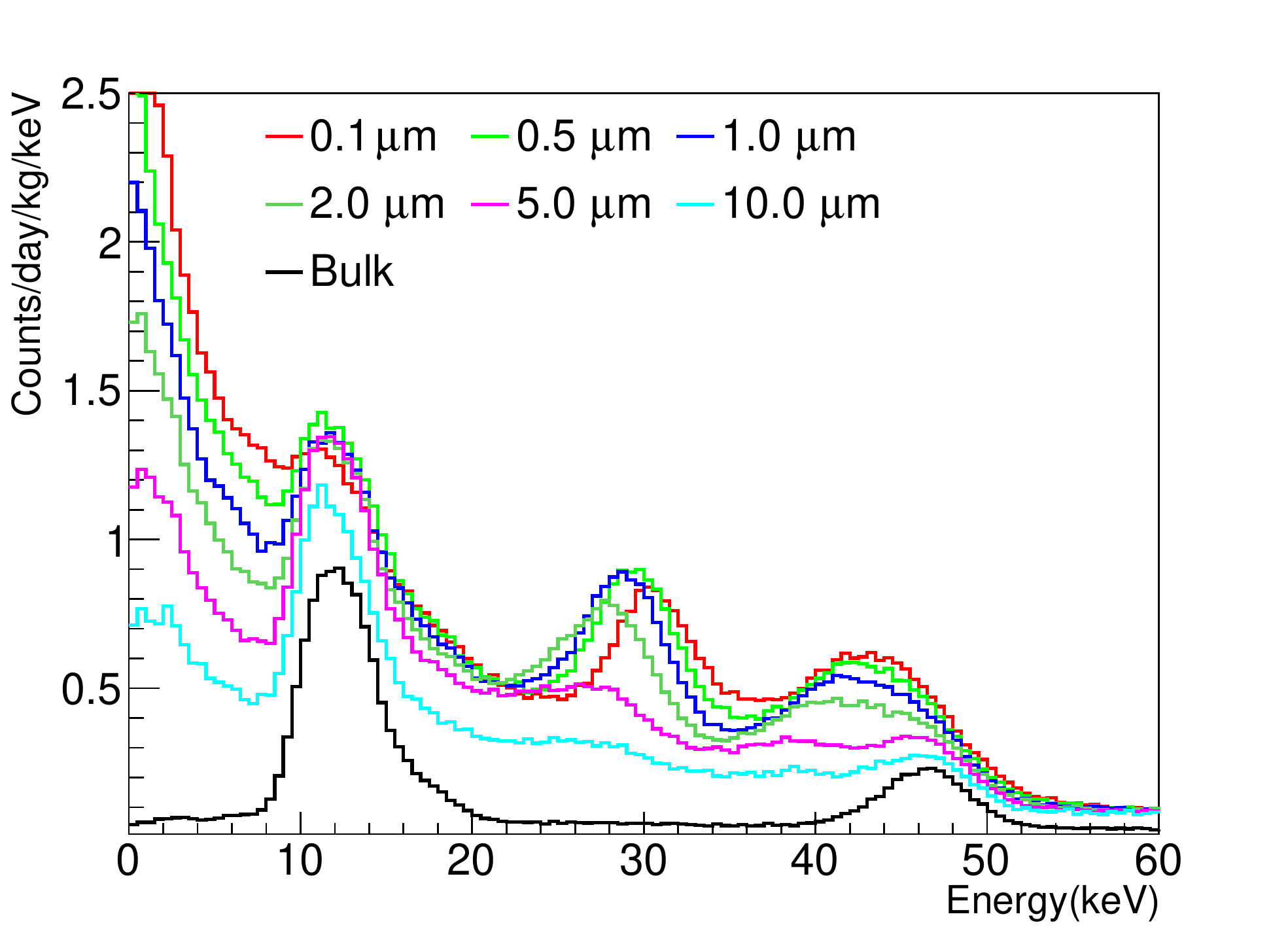}
\caption{
Comparison of background spectra of $\rm ^{210}Pb$ simulated for various surface thicknesses of PTFE reflector. The activity of 1~mBq/kg is used to normalize the simulation results.
}
\label{C3_tef_Pb210}
\end{center}
\end{figure}

\begin{figure*}
\begin{center}
\begin{tabular}{cc}
\includegraphics[width=0.495\textwidth]{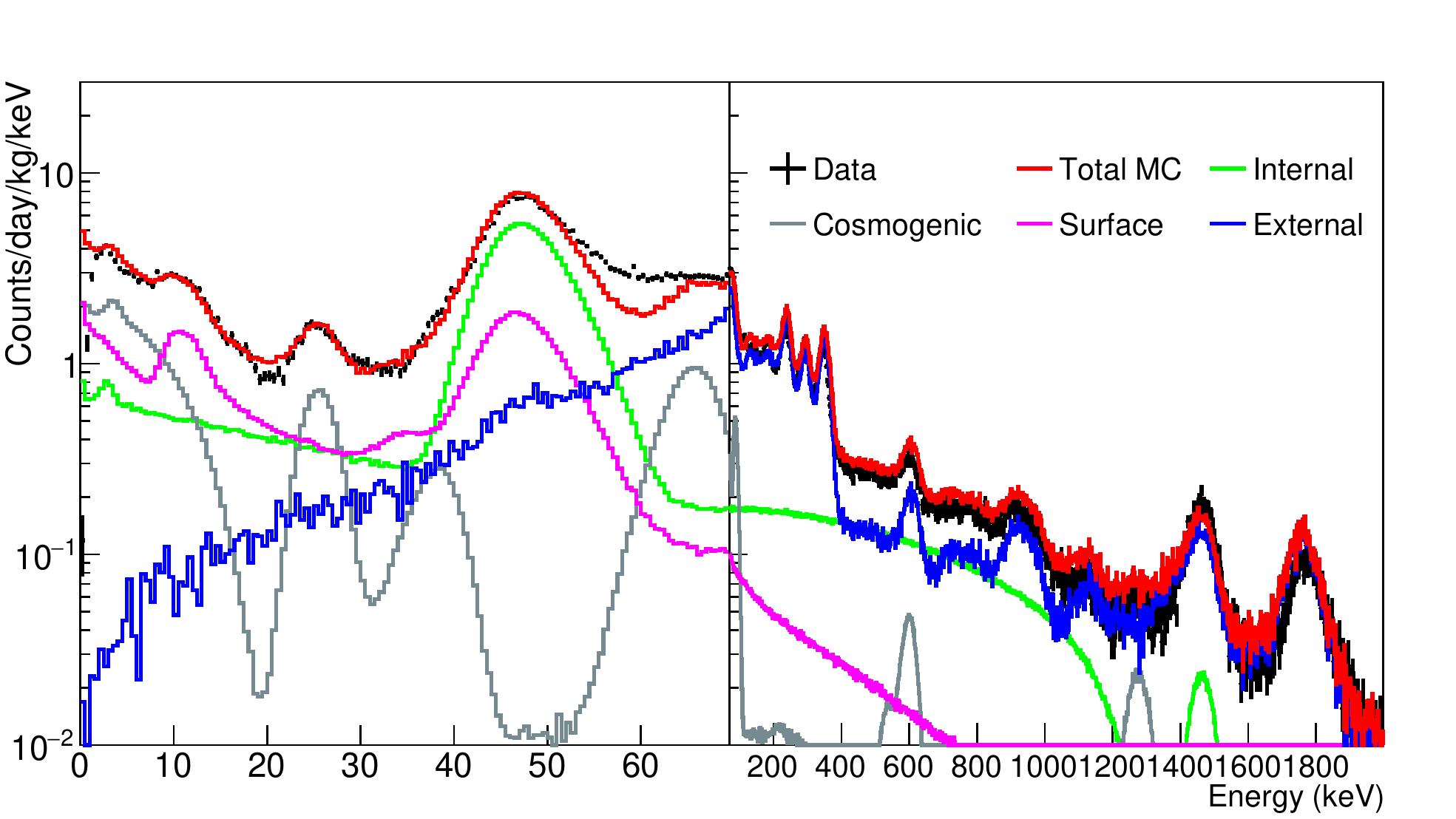}  &
\includegraphics[width=0.495\textwidth]{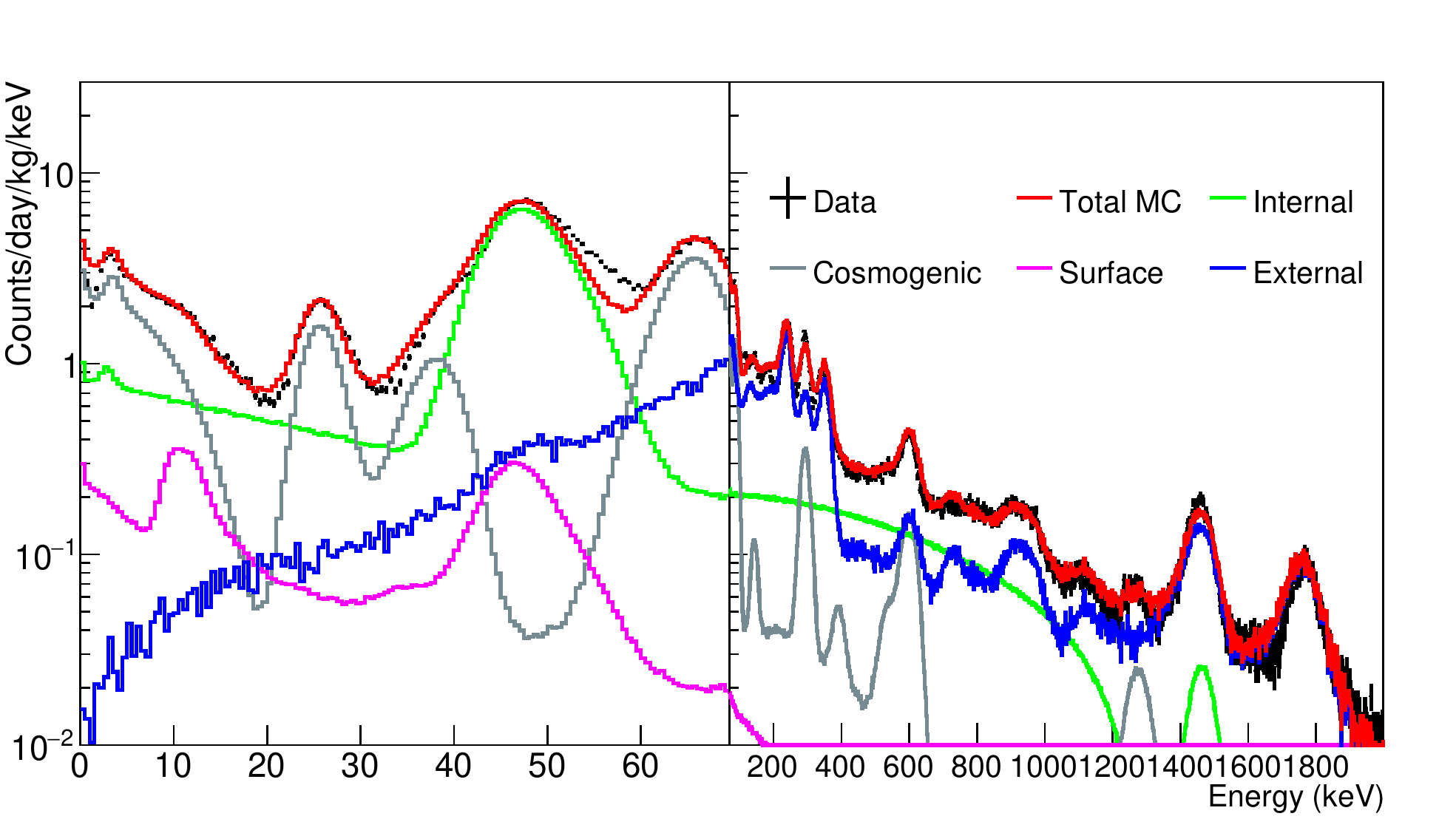} \\
(a) Crystal-3 & (b) Crystal-4
\end{tabular}
\caption[]{Measured single-hit background spectra of Crystal-3 and Crystal-4 fitted with all simulated background spectra.}
\label{C3_fit}
\end{center}
\end{figure*}
%
%\begin{figure*}
%\begin{center}
%\includegraphics[width=0.75\textwidth]{C4_comparison_all_single_2.pdf}
%\caption[]{Measured single-hit background spectra of Crystal-4 fitted with all simulated background spectra.}
%\label{C4_fit}
%\end{center}
%\end{figure*}

Although Crystal-3 and 4 were both grown with WIMP-Scint-II grade powder by Alpha Spectra in July 
2014, they were delivered to the Y2L at different times.
Cryst-al-3 has a complicated exposure history and was repaired once before arriving at Y2L in July 2015 and has remained underground ever since. On the other hand, Crystal-4 was delivered at the Y2L in March 2016 after being exposed to surface-level cosmic rays for about 2 years and was installed for COSINE-100 after just six months of cooling.
As a result, Crystal-4 is expected to have short-lived cosmogenic isotopes as well as long-lived cosmogenic isotopes.

The background spectrum of Crystal-3, shown in Fig.~\ref{C3_fit}a, has a peak around 10~keV that has not changed over time. 
To understand its origin, we studied the effect of surface $^{210}$Pb from PTFE reflective sheets that wrapped each crystal's lateral surfaces in ten layers with 250~$\mu$m total thickness. We simulated the background spectrum of $^{210}$Pb by generating it randomly %on the surface of the PTFE sheets for various depths
within the layer of the PTFE sheets with various thicknesses: 0.1, 0.5, 1.0, 2.0, 3.0, 5.0~$\mu$m, and also in the bulk. The simulated spectra are shown in Fig.~\ref{C3_tef_Pb210}, where each color represents the different surface depths and the bulk reflector (black solid line). 
The peaks at 12~keV and 46~keV, which are prominent for the bulk reflector, are due to the X-rays and 46.5~keV $\gamma$-ray from the decays of $^{210}$Pb, respectively. 
Conversion electrons contribute peaks around 30~keV and 43~keV and beta electrons contribute a continuum at peaks at low energy.
Since the conversion electrons' energy losses depend on the thickness of PTFE that they penetrate, these peaks move to lower energies as the depth increases.
In the simulation we used the spectrum of $^{210}$Pb from the bulk reflector to model the 12~keV peak because the surfaces can be treated as a bulk with $\sim$ 10 layers of PTFE sheets. 

The Crystal-4 spectrum, shown in Fig.~\ref{C3_fit}b, has three correlated peaks from the decay of $^{109}$Cd: $\sim$3~keV and $\sim$25~keV %X-rays 
binding energies from L-shell/K-shell electron captures and $\sim$88~keV gamma-ray line from the isomer transition of $^{109}$Ag. We also determined its half-life by measuring the change of $\sim$25~keV %X-ray
binding energies contribution over time with a result that is in a good agreement with the 462 day expectation.
The activity from the fit to the data, found to be 0.09$\pm$0.06~mBq/kg,  is consistent with the measurement. 

\subsection{Background model for crystals 6 and 7}
\label{sec:4.3}

\begin{figure*}
\begin{center}
\begin{tabular}{cc}
\includegraphics[width=0.495\textwidth]{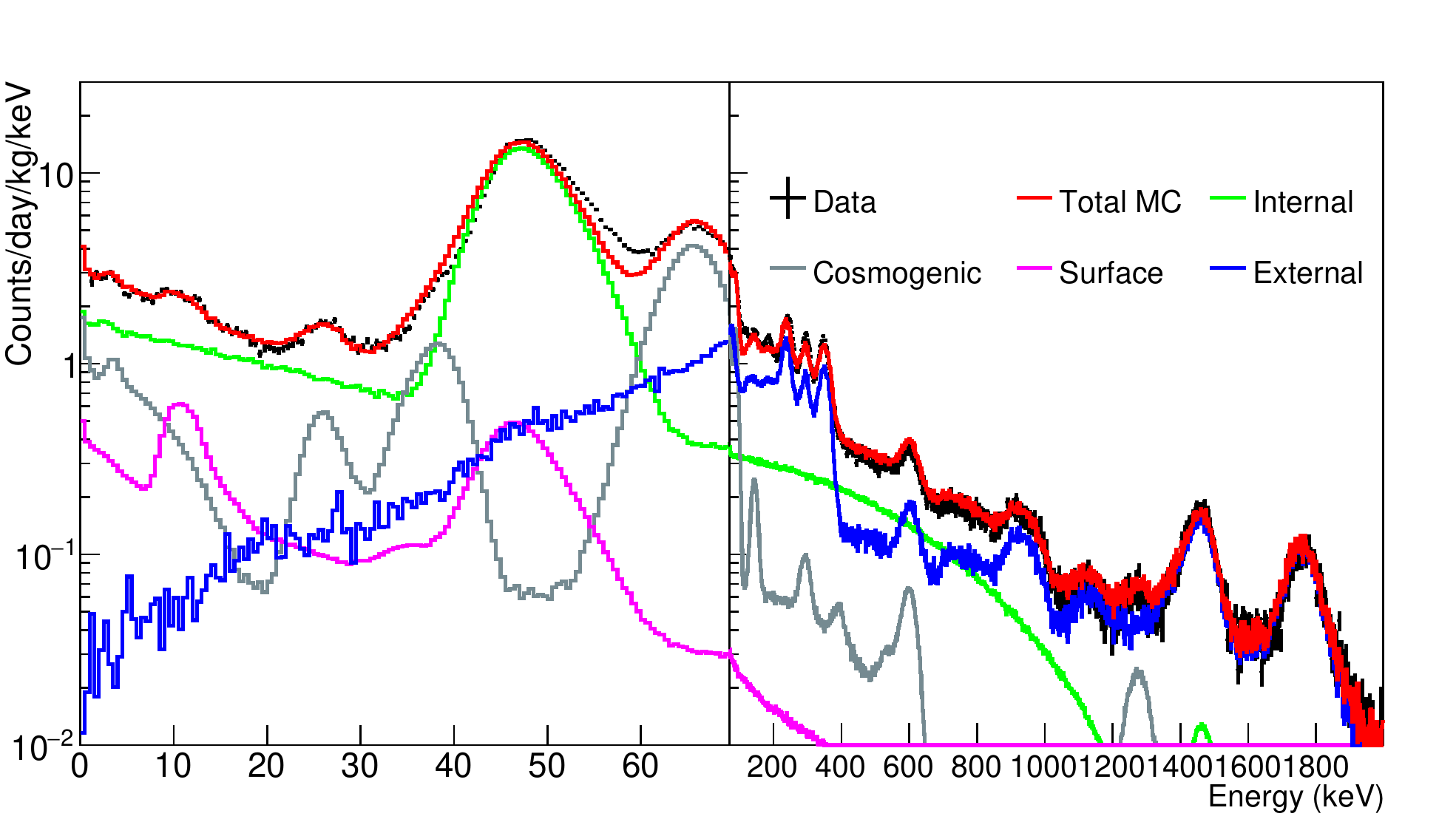} &
\includegraphics[width=0.495\textwidth]{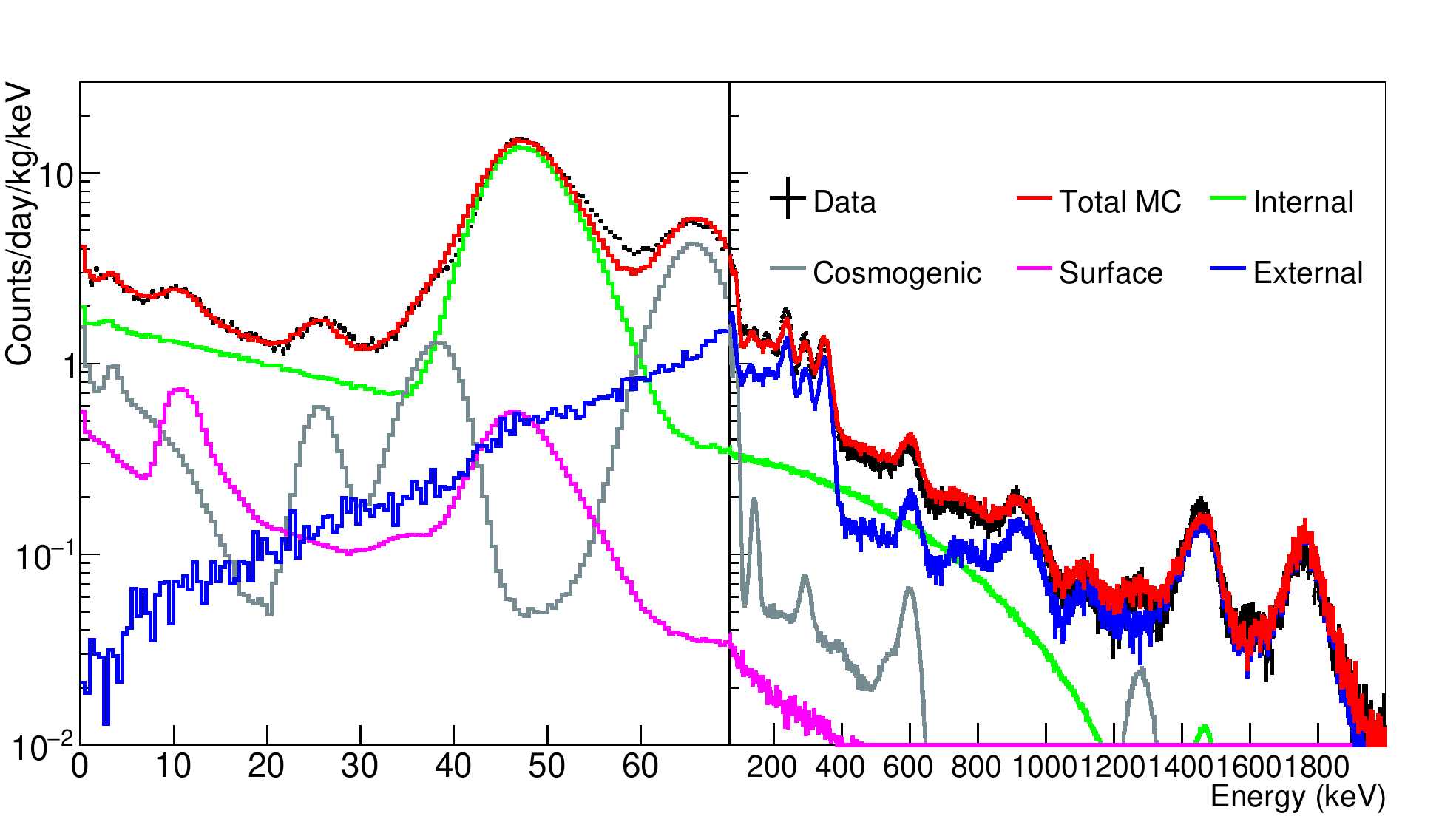} \\
(a) Crystal-6  & (b) Crystal-7
\end{tabular}
\caption[]{Measured single-hit background spectra of Crystal-6 and Crystal-7 fitted with all simulated background spectra.}
\label{C67_fit}
\end{center}
\end{figure*}

Crystal-6 and 7 are twin crystals made from WIMPSci-nt-III grade powder by Alpha Spectra at the same time. 
They were installed 7 months after their delivery to Y2L, similar to Crystal-4. However their surface exposure times were shorter than those of the other crystals.
Figure~\ref{C67_fit} shows that the fitted simulation spectra accurately reproduce the measured data.
As expected, they have similar contamination levels of short-lived cosmogenic isotopes.

\section{Discussion of fitted background spectra}
\label{sec:5}

%\makeatletter
%\DeclareRobustCommand{\textsupsub}[2]{{%
%  \m@th\ensuremath{%
%    ^{\mbox{\fontsize\sf@size\z@#1}}%
%    _{\mbox{\fontsize\sf@size\z@#2}}%
%  }%
%}}
\renewcommand{\arraystretch}{2}
\renewcommand{\arraystretch}{2}

We have simulated background spectra from radioactive sources: full decay chains of $^{238}$U, $^{232}$Th, and $^{40}$K from the crystals, 16 PMTs, and the other external sources; bulk and surface $^{210}$Pb from the crystals; surface $^{210}$Pb from PTFE reflector; $^{125}$I, $^{121}$Te, $^{121m}$Te, $^{123m}$Te, $^{125m}$Te, $^{127m}$Te, $^{113}$Sn, $^{109}$Cd, $^{22}$Na, and $^{3}$H for cosmogenic isotopes,   
and have fitted them to the data to estimate their unknown contamination levels.

\begin{figure*}
\begin{center}
\begin{tabular}{cc}
\includegraphics[width=0.48\textwidth]{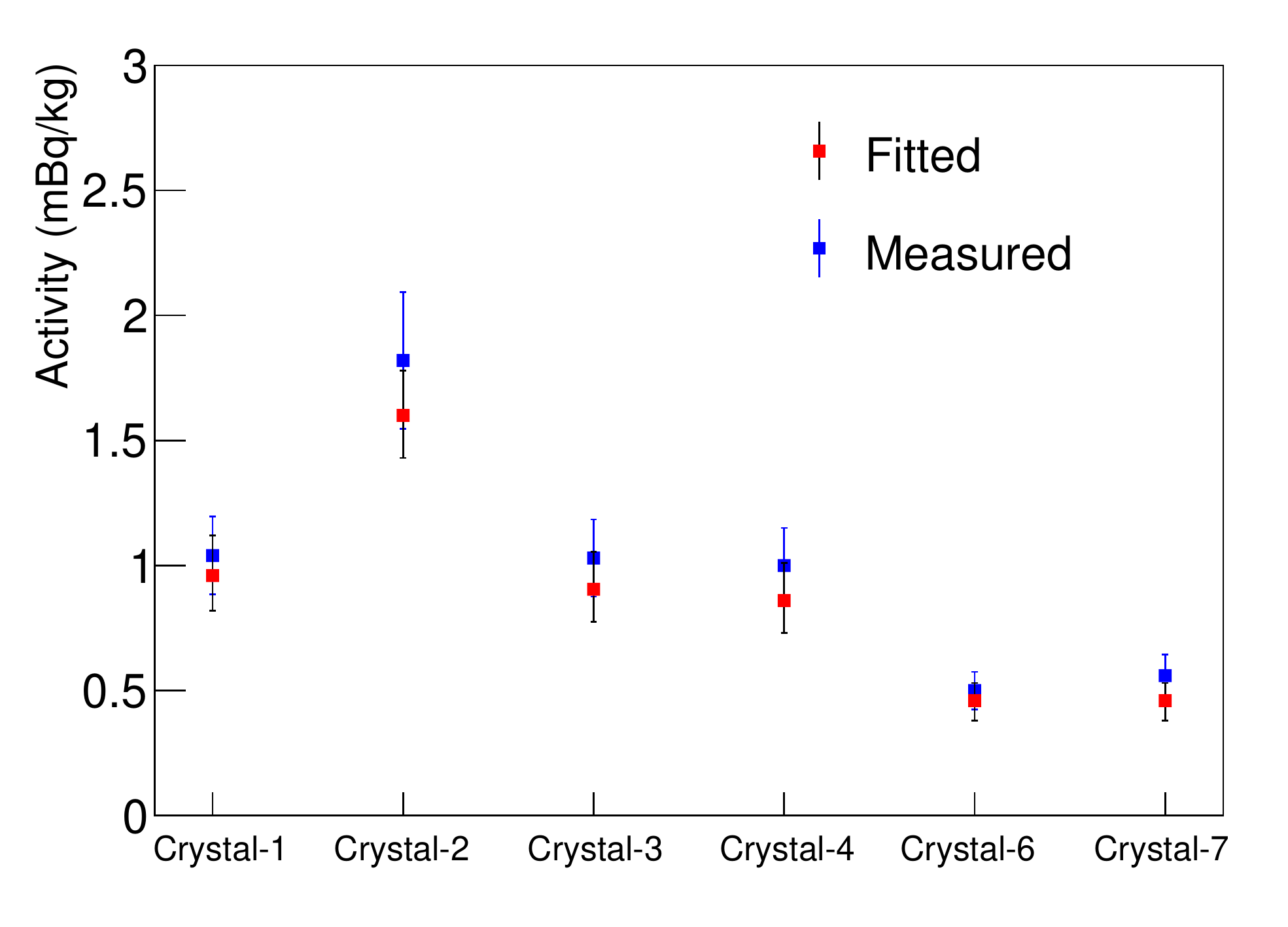} &
\includegraphics[width=0.48\textwidth]{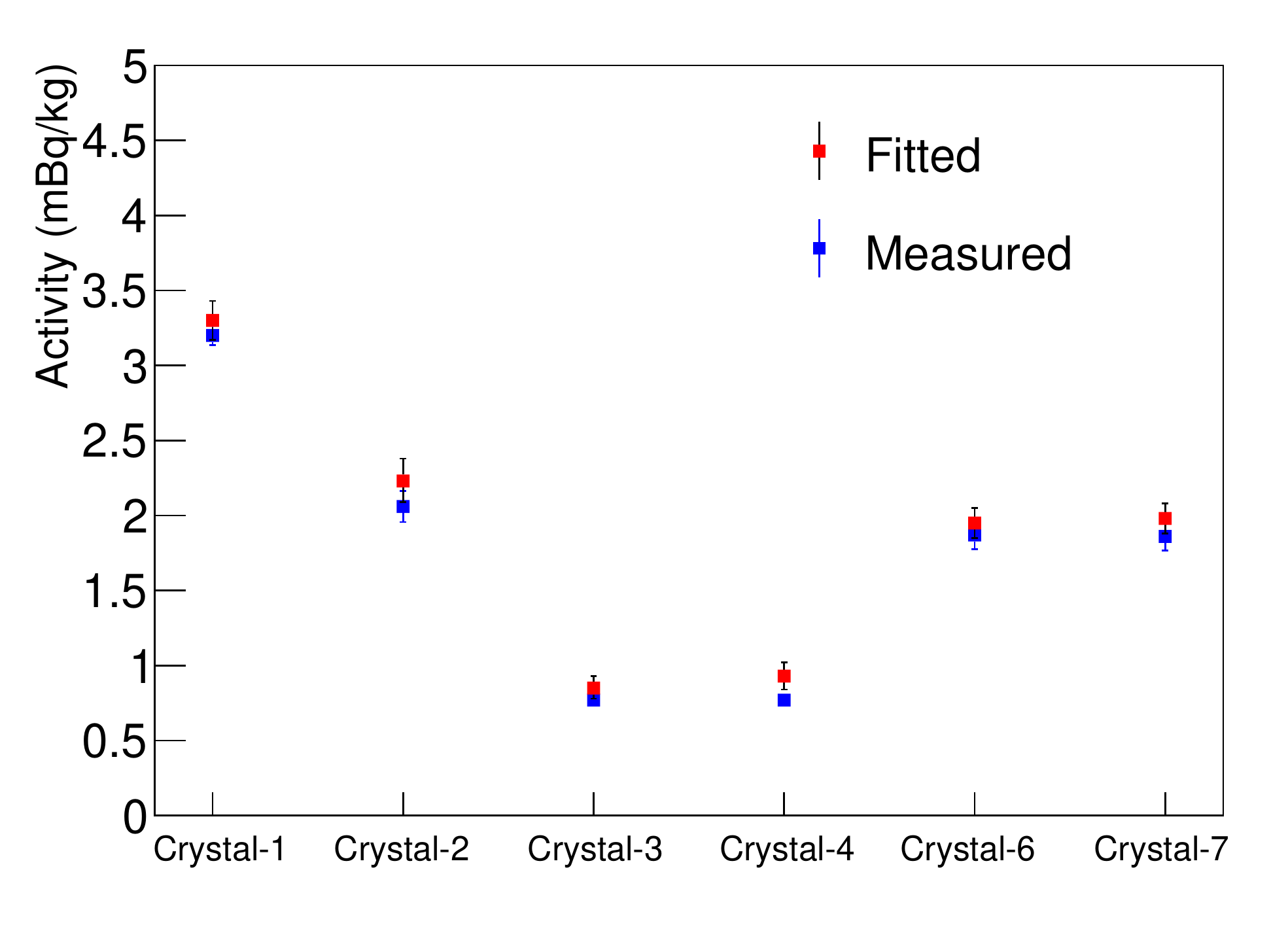} \\
(a) Internal $^{40}$K & (b) Bulk $^{210}$Pb \\
\end{tabular}
\caption[]{Comparison of the measured and the fitted activity levels in six NaI(Tl) crystals. 
}
\label{Comparison_Act}
\end{center}
\end{figure*}

\begin{figure}
\begin{center}
\includegraphics[width=0.48\textwidth]{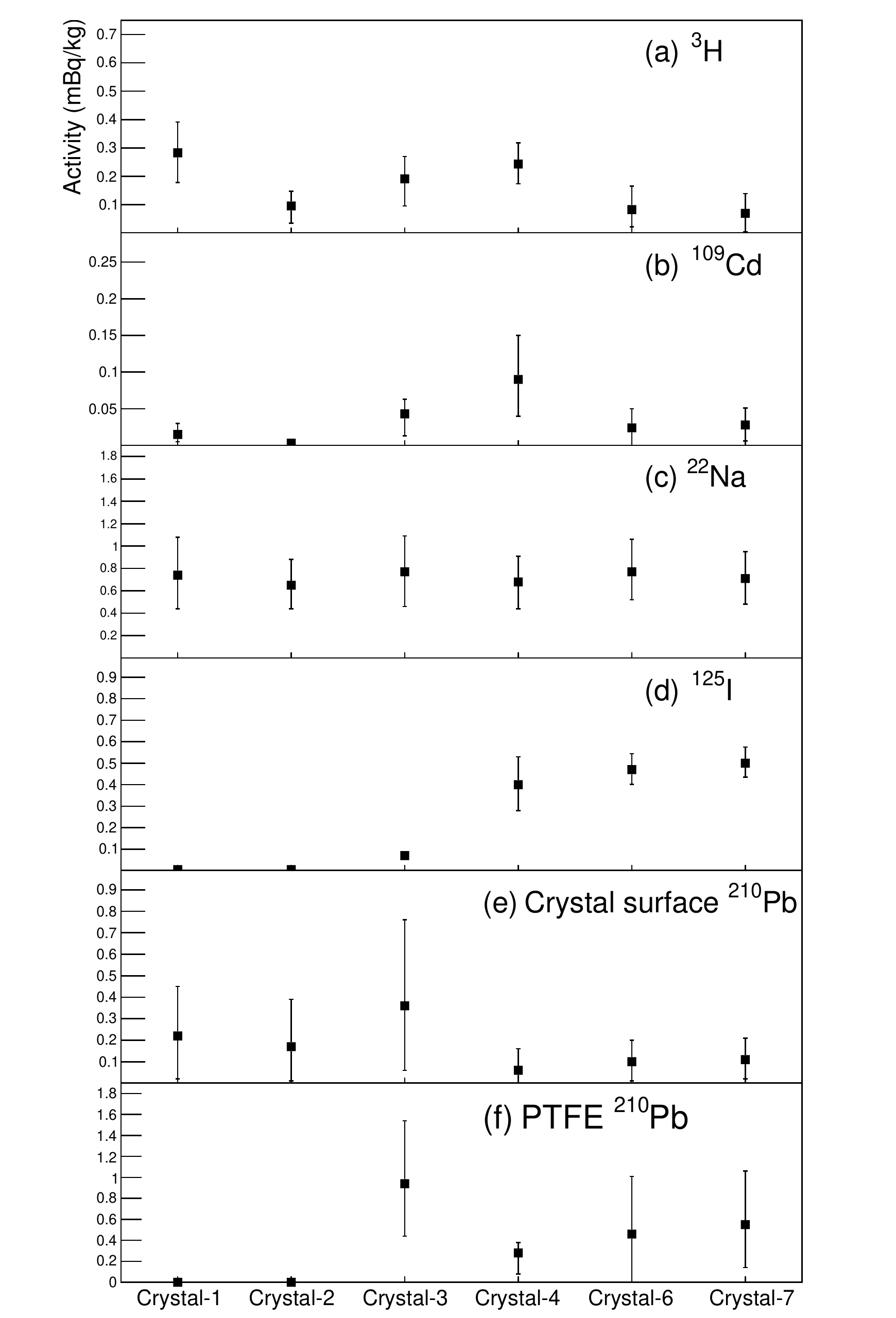}
\caption{
Fitted activities for (a) $^{3}$H, (b) $^{109}$Cd, (c) $^{22}$Na, (d) $^{125}$I,  (e) Crystal surface $^{210}$Pb, and (f) PTFE $^{210}$Pb that correspond to the averaged activities between Oct 21 and Dec 19 2016. Activities of crystal surface $^{210}$Pb and PTFE $^{210}$Pb were normalized by a total mass of the crystal.
}
\label{fittedresults}
\end{center}
\end{figure}

In Fig.~\ref{Comparison_Act} we compare the fitted activities of internal $^{40}$K and $^{210}$Pb to their measured levels for the six crystals, where there is agreement at the $\sim$20\% level. 
The fitted activities of cosmogenic isotopes: $^{3}$H, $^{109}$Cd, $^{22}$Na, and $^{125}$I, surface $^{210}$Pb, and PTFE $^{210}$Pb for the six crystals are shown in Fig.~\ref{fittedresults}.
As explained in section~\ref{sec:4}, Crystal-1 and 4 have two-year-long exposure times and, thus, relatively high $^{3}$H levels (see Fig.~\ref{fittedresults}(a)). 
The largest contribution of $^{109}$Cd is in Crystal-4 (Fig. \ref{fittedresults}(b)) and the fitted activity is in a good agreement with the measured value.
The fitted activities of $^{125}$I in Crystal-4, 6, and 7 are similar to each other and higher than other crystals due to their relatively short cooling times underground, as shown in Fig.~\ref{fittedresults}(c).
There are more contributions from surface $^{210}$Pb in Crystal-1, 2, and 3,
and small contributions from PTFE $^{210}$Pb in Crystal-1 and 2. 
Crystal-3 has large contributions from the surface $^{210}$Pb for both the crystal itself and the PTFE reflector, albeit with large errors.

\begin{figure}
\begin{center}
\includegraphics[width=0.5\textwidth]{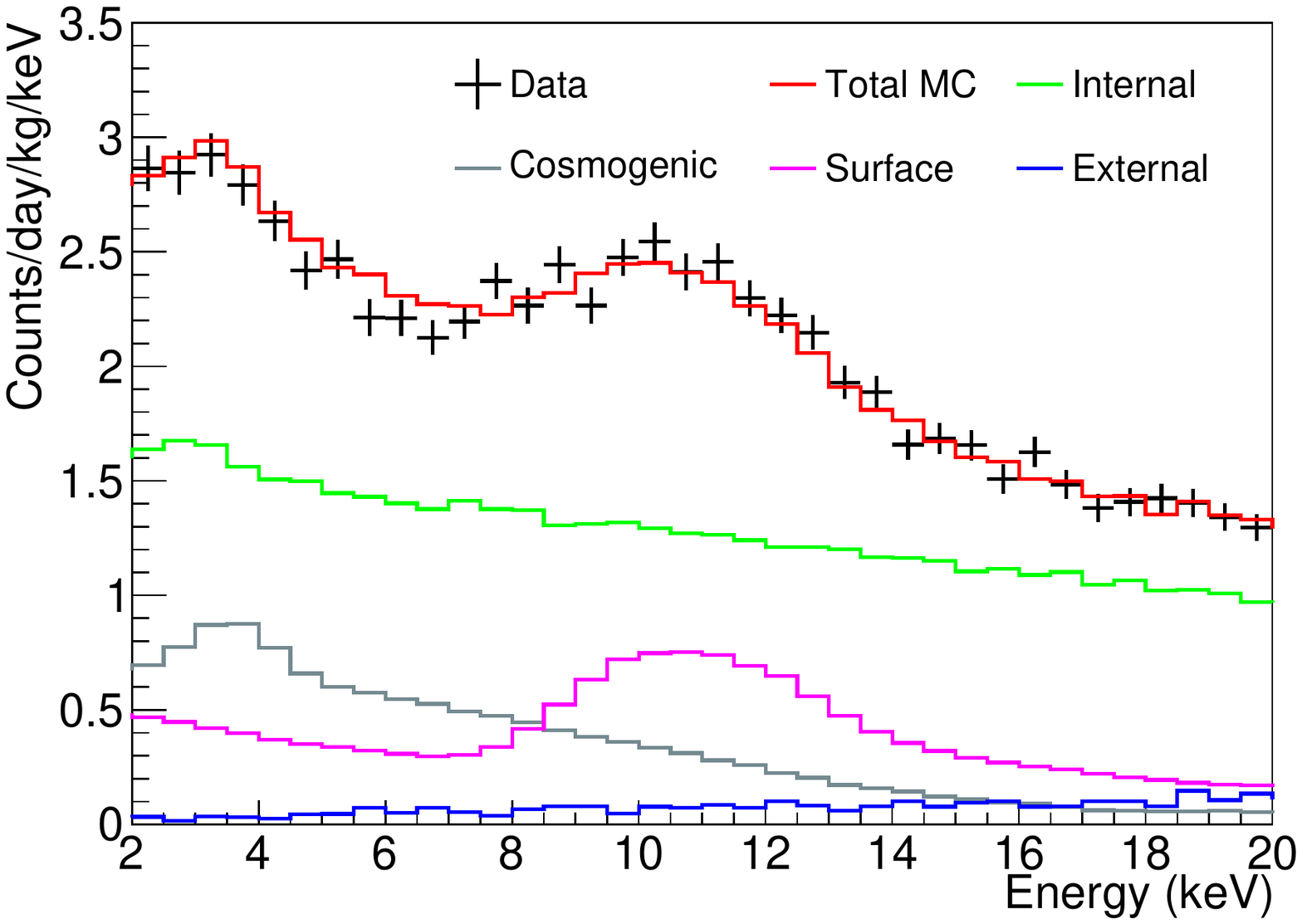} \\
\caption[]{
Comparison of data and Monte Carlo simulation of the low-energy background spectra of Crystal-7.
}
\label{C6_zoom}
\end{center}
\end{figure}

Based on the background model for the six NaI(Tl) detectors 
we determine the background levels in units of dru (counts/day/keV/kg) in the 2-6~keV energy interval and list them in Table~\ref{Tab_low}.
The dominant background contributions are from $\rm ^{210}Pb$ and $\rm ^{3}H$; representative background spectra of Crystal-7 in the 2-6~keV energy region are shown in Fig.~\ref{C6_zoom}. 
The overall simulated background spectra well describe the measured data for the six NaI(Tl) detectors. 

\begin{table*}
\begin{center}
\caption{
Fitted background events in units of dru (counts/day/keV/kg) in the (2--6) keV energy interval.
}
\label{Tab_low}
\begin{tabular}{lccccccc}
            &     &Crystal-1   & Crystal-2  & Crystal-3   & Crystal-4    & Crystal-6   & Crystal-7\\
\hline
 Internal  & $\rm ^{40}K$~ &0.10  $\pm$0.02 & 0.20  $\pm$ 0.02   &0.10  $\pm$ 0.01    & 0.10 $\pm$0.01   &0.05    $\pm$ 0.01   & 0.05 $\pm$0.01  \\
               &  $\rm ^{210}Pb$~ & 2.50 $\pm$ 0.10 &1.69 $\pm$ 0.09  &0.57 $\pm$ 0.05  &0.71 $\pm$ 0.05  & 1.46  $\pm$  0.07 &1.50 $\pm$ 0.07 \\
                & Other ($\times10^{-4}$)  & 7.0$\pm$0.1 &  15$\pm$1 & 7.3$\pm$0.1 & 7.7$\pm$0.1  & 14$\pm$1  & 14$\pm$1  \\
 Cosmogenic  &$\rm ^{3}H$ & 2.35 $\pm$ 0.90  &0.81 $\pm$ 0.40   &1.54 $\pm$ 0.77  &1.97 $\pm$ 0.66  & 0.69  $\pm$ 0.67  &0.58 $\pm$ 0.54  \\
             &$\rm ^{109}Cd$ & 0.05 $\pm$0.04  &0.009 $\pm$0.009   &0.13 $\pm$ 0.06   &0.29 $\pm$ 0.15  & 0.08  $\pm$ 0.08  &0.09 $\pm$ 0.09 \\
             & Other & -  & -  &0.02 $\pm$ 0.01  &0.09 $\pm$ 0.04  & 0.06  $\pm$  0.03 &0.05 $\pm$ 0.03 \\

 Surface & $^{210}$Pb    & 0.64 $\pm$ 0.64   &0.51 $\pm$ 0.51  & 1.16 $\pm$ 0.51 & 0.22 $\pm$ 0.16 &0.34  $\pm$  0.20  &0.38 $\pm$ 0.21 \\
 External  & & 0.03 $\pm$ 0.02  &0.05 $\pm$ 0.04  &0.03 $\pm$ 0.02  & 0.03 $\pm$ 0.02 &0.04  $\pm$  0.03  &0.03 $\pm$ 0.02 \\

 Total simulation & &5.68 $\pm$ 1.04  & 3.28 $\pm$ 0.67 &3.57 $\pm$ 0.76  & 3.41 $\pm$ 0.75 & 2.74  $\pm$  0.61 &2.70 $\pm$ 0.51 \\
 Data  & &5.64 $\pm$ 0.10  &3.27 $\pm$ 0.07  & 3.35 $\pm$ 0.07 &3.19 $\pm$ 0.05  &2.62  $\pm$  0.05  &2.64 $\pm$ 0.05 \\
\hline
\end{tabular}
\end{center}
\end{table*}

\section{Conclusion} 
\label{conc}
%We have studied background levels using the Geant4 toolkit, of the NaI(Tl) crystal detectors 
We have studied, using the Geant4 toolkit, the background of the NaI(Tl) crystal detectors
that are being used in the COSINE-100 dark matter search experiment. 
The crystals have different exposure histories and underground radioactivity cooling times. 

In the background modeling the overall energy spectrum summed over all simulations is well matched to the data not only for single-hit events but also for multi-ple-hit events.
Crystal-1 and Crystal-2 that had cooling times as long as three years at the Y2L are dominated by $^{210}$Pb and $^{3}$H for energies below 20~keV.
The background contribution of $^{3}$H in Crystal-2 is smaller than that in Crystal-1 due to its shorter surface exposure time.
Crystal-6 and 7 show clear contributions from $^{125}$I due to their short cooling times underground, as expected. 
Crystal-3 %has a complicated exposure history 
had an additional treatment for a repair that increased the background near 10~keV that is well modeled by surface $^{210}$Pb on the PTFE wrapping foil. 
Crystal-4 was exposed to surface cosmic rays for two years and only had a six month-long underground cooling time. As a result, this crystal has significant background contributions from both short-lived and long-lived cosmogenic isotopes.

Background contributions from external sources and internal $^{40}$K are reduced to the level of 0.03 dru and about 0.1 dru in the energy range of 2--6~keV, respectively, by the LS veto detector that surrounds the crystals. 

The average background rate in the (2-6) keV energy range for the six crystals (with a total mass of 70 kg) studied here is 3.5 counts/day/keV/kg.  
The dominant contributions in this energy range are from $^{210}$Pb and $^3$H.
   
\section*{Acknowledgments}
We thank the Korea Hydro and Nuclear Power (KHNP) Company for providing underground laboratory space at Yangyang.
This work is supported by: the Institute for Basic Science (IBS) under project code IBS-R016-A1, Republic of Korea; UIUC campus research board, the Alfred P. Sloan Foundation Fellowship, NSF Grants no. PHY-1151795, PHY-1457995, DGE-1122492 and DGE-1256259, WIPAC, the Wisconsin Alumni Research Foundation, Yale University and DOE/NNSA Grant no. DE-FC52-08NA28752, United States; STFC Grant ST/N000277/1 and ST/K001337/1, United Kingdom; and CNPq and Grant no. 2017/02952-0 FAPESP, Brazil.

%
% BibTeX users please use
% \bibliographystyle{}
% \bibliography{}

\end{document}